**Prevalent Intrinsic Emission from Nonaromatic Amino Acids and Poly(Amino Acids)**

*Xiaohong Chen, Weijian Luo, Huili Ma, Qian Peng, Wang Zhang Yuan\*, and Yongming Zhang\**

X. Chen, W. Luo, Prof. W. Z. Yuan, Prof. Y. Zhang
School of Chemistry and Chemical Engineering
Shanghai Key Lab of Electrical Insulation and Thermal Aging
Shanghai Electrochemical Energy Devices Research Center
Shanghai Jiao Tong University
No. 800 Dongchuan Rd., Minhang District, Shanghai 200240, China
E-mail: wzhyuan@sjtu.edu.cn; ymzsjtu@gmail.com
Dr. H. Ma
Department of Chemistry, Tsinghua University
Zhongguancun Street, Haidian District, Beijing 100084, China
Prof. Q. Peng
Institute of Chemistry, Chinese Academy of Sciences
Zhongguancun North First Street 2, Haidian District, Beijing 100190, China

Intrinsic luminescence of naturally occurring proteins provides considerable information about their structure and dynamics and thus is often used to study protein folding, associated reaction, and other biochemical processes.[1,2] Generally, protein emission is believed to be originated from three aromatic amino acids, namely phenylalanine (Phe), tyrosine (Tyr), and tryptophan (Trp).[2,3] Little attention, however, has been paid to nonaromatic amino acids, presumably due to their lack of classic discrete fluorophores. Recently, several groups reported the intrinsic emission form the aggregates of some biomolecules.[4-7] Discovery of the bright intrinsic emission of peptides and proteins allows for direct imaging of their folding and aggregation properties.[6a,7a] In particular, the study of the intrinsic fluorescence from proteins permits sensitive information about the native protein to be obtained with little or no structural modification, thus highly valuable in biochemical researches.[7a] Despite exciting advances have been achieved, there still deserves more insightful understanding on the emission mechanism, for which different and even controversial hypotheses were proposed.

For example, Homchaudhuri and coworkers observed the emission from concentrated solutions of *L*-lysine monohydrochloride,[4a] and ascribed it to the gathering of amino (NH$_2$) side chains through further investigation of poly(*L*-lysine).[4b] Shukla et al. serendipitously discovered intrinsic emission from different nonaromatic protein crystals and aggregates.[5] They hypothesized that delocalization of peptide electrons by extensive arrays of intra- and/or intermolecular hydrogen bond formation.[5,6b,6c] However, de Mercato et al. suggested hydrogen-bonded water molecules within the cross-*β* structure are responsible for the peptide emission.[6d] And very recently, Ye et al. ascribed the peptide emission to the communication of amide groups.[7b]

Apart from above biomolecules, previously, other nonconventional luminogens[8-12] like natural products,[8a] polyacrylonitrile (PAN),[8b] polyamidoamines (PAMAM),[9] and polyureas (PU)[10] were also reported, accompanying with diverse assumptions for the emission.[8,9,11a] In 2013, based on the observation of amazing emission of rice, starch, and cellulose, we proposed the clustering of electron rich oxygen units and the subsequent electron cloud overlap (delocalization) and conformation rigidification to rationalize the emission.[8a] Such clustering-triggered emission (CTE) mechanism was further supported by additional facts in the PAN system.[8b] Although the genesis of the emission is not yet settled, a fair amount of data has accumulated to suggest that CTE mechanism, namely the clustering of nonconventional chromophores and subsequent electron cloud overlap, is not only reasonable to account for our systems, but also applicable to rationalize other systems.[8c]

The intriguing emission from both nonaromatic biomolecules and the other nonconventional natural or synthetic luminogens inspirited us to think the general fundamental principle underlying the phenomena. In this contribution, to acquire more insights into the emission of biomolecules, and to unveil more common grounds and correlations for both general and biomolecular nonconventional luminogens, we thoroughly investigated the photophysical properties of the most basic building blocks of biomolecules, namely, nonaromatic amino acids. Compared with the intensive investigations of aromatic amino acids, their emission behaviors, particularly solid-state emission remains virtually unexplored. Meanwhile, poly(amino acids), for which *ε*-poly-*L*-lysine (*ε*-PLL) was chosen as an representative, were also checked. Newly revealed facts illustrate the widespread occurrence of intrinsic emission from these biomolecules. As example in **Figure 1**, nonaromatic amino acids with different side chains, including simple alkane group, hydroxyl (OH), sulfur (SH), carbonyl (C=O), and $NH_2$ containing units, can generate noticeable visible light emission even under 365 nm UV irradiation. Furthermore, besides shortlived fluorescence, longlived and even persistent room temperature phosphorescence (RTP) are also detected, which may reminder us to reevaluate the origin of RTP from proteins, particularly at the solid states. Taken together with the emission behaviors and single crystal analysis, CTE mechanism is adopted to rationalize the intrinsic visible emission. These results should be highly implicative for further understanding on the emission of biomolecules at varying states.

Whilst dilute aqueous solutions of nonaromatic amino acids are nonemissive, noticeable visible emissions are observed in the majority of their concentrated counterparts and all of the recrystallized solids with efficiencies (*Φ*) up to 7.4% (**Figure** 1, **2**, and S1–S9, Supporting Information),[13] demonstrating concentration enhanced emission and aggregation-induced emission (AIE) characteristics.[8,14] Above observations are somewhat different from previous reports, wherein the observed luminescence is unique to *L*-Lys hydrochloride salt alone, and the randomly chosen glycine (Gly), *L*-arginine (*L*-Arg), *L*-serine (*L*-Ser), *L*-glutamate (*L*-Glu), and *L*-isoleucine (*L*-Ile) are nonluminescent.[4a] Such distinction is highly possibly caused by the variation of sample states. Therefore, to ascertain the genesis of the intrinsic visible emission of these compounds, it is important to thoroughly investigate their photophysical properties at varying states. Herein, *L*-Lys, *L*-Ser, and *L*-Ile were chosen as the examples to show the general aspects of various nonaromatic amino acids. Negligible photoluminescence (PL) is observed for their dilute solutions, whereas considerable PL rise is noticed for their concentrated counterparts (Figure 2, S8, and S9, Supporting Information). The $\Phi$ values for the dilute ($8\times10^{-5}/10^{-3}/10^{-3}$ M) and concentrated (0.1/0.5/0.2 M) solutions of *L*-Lys/*L*-Ser/*L*-Ile are approaching zero and 7.0%/2.2%/2.3%, respectively. This trend is consistent with the direct observation. Notably, for *L*-Lys, when the concentration is above 0.1 M, concentration quenching occurs, which should be ascribed to the self-absorption and/or exciton coupling. With an excitation wavelength ($\lambda_{ex}$) of 365 nm, emission maxima at 440/488, 438/493, and 438/483 nm are monitored for *L*-Lys (0.1 M), *L*-Ser (0.5 M), and *L*-Ile (0.2 M), with lifetimes ($<\tau>$) of 4.90/5.06, 2.82/4.07, and 2.67/3.96 ns (Figure 2, S8–S10, Supporting Information), respectively, thus indicating the existence of multiple emissive species. Such heterogeneous population is also supported by their $\lambda_{ex}$-dependent emission. Clearly, the emission maxima of concentrated solutions of *L*-Lys/*L*-Ser/*L*-Ile vary from 433/427/422 to 511/529/492 nm while their $\lambda_{ex}$s being changed from 340/312/310 to 460/480/420 nm (Figure 2C, S8C, and 9C, Supporting Information).

Absorption spectra were further recorded to gain more information (Figure 2B, S11–S14, Supporting Information). For *L*-Lys, its dilute solutions give rather weak absorption, which is enhanced with increasing concentration. And apparent shoulders around 273 and 328 nm are detected when the concentration is no less than $2\times10^{-3}$ M. Meanwhile, a gradual extension of absorption edge with increasing concentration is monitored, which even enters into the region above 500 nm ($\geq$0.25 M). Such extended absorption might be accountable for the decreased emission owing to energy transfer. Taken together the studies above establish a likelihood for the presence of aggregates in concentrated solutions. This conjecture is supported by the dynamic light scattering (DLS) measurement, which suggests the formation of nanoaggregates with sizes ranging from ~40 to 350 nm (Figure S15, Supporting Information). Cryogenic experiment further discloses the nonluminescence of the dilute solutions even at 77 K (Figure 2E, S8D, and S9D, Supporting Information), which excludes conformation rigidification as the sole cause for the emission. However, emissions of concentrated solutions are greatly promoted due to further conformation rigidification. Moreover, persistent phosphorescence from the solid glass after ceasing UV illumination is observed (Figure 2E). These results may have significant implications for the protein phosphorescence, which is previously believed to be exclusively stemmed from aromatic amino acids.[3]

Above emission behaviors are similar to those of PAN,[8b] highly indicative of the strong correlation of their underlying mechanism. Based on these experimental results, such unique emission of nonaromatic amino acids can be well rationalized by the CTE mechanism.[8] In dilute solutions, these compounds are molecularly dissolved as individuals, which are difficult to be excited owing to their insufficient conjugation. In the concentrated solutions, however, they may approach each other with the aid of intermolecular interactions like hydrogen bonds, forming diverse nanoaggregates. Albeit there is no classic aromatic conjugation, the presence of amino, carbonyl, and hydroxyl subunits affords further through space electronic communications between π electrons and lone pairs (n), resulting in electron cloud overlap and thus extended electron delocalization together with simultaneously rigidified conformations. Subsequently, these clustered chromophores can be readily excited even with UV-A irradiation to yield visible emission, which can be boosted upon further conformation rigidification.

Apart from these exampled amino acids, other nonaromatic systems and their mixtures (exampled by *L*-Ile-*L*-Arg and *L*-Ile-Gly) also demonstrate resembling photophysical properties (Figure S1–S6, and S16, Supporting Information), thus confirming the general occurrence of intrinsic emission. To gain more insights into the mechanism, their photophysical properties at solid state were studied (**Figure 3**, S17–S26, Supporting Information). When excited with different $\lambda_{ex}$s, $\lambda_{ex}$-dependent emission profiles with relatively minor variations in emission peaks/shoulders are recorded for the recrystallized samples (Figure 3). And similar peak positions around 385, 422, 484, 530, and 598 nm are identified for different compounds, indicative of resembling conjugations in diverse systems. With comparison to solution emissions, those of crystalline solids display much narrower full width at half maximum (FWHM), which is presumably ascribed to the conformation rigidification or quantum well structure of the crystals.[15] Such multiple emission maxima should be stemmed from diverse emissive species, as confirmed by their different <$\tau$> values (Figure S19, Supporting Information).[16] Meanwhile, distinctive emission colors are recorded for *L*-Lys solids under varying excitations (Figure 3D), suggestive of the heterogeneous population of excited states. Specifically, with a delay time ($t_d$) of 0.1 ms,[17] triplet emissions peaking at 524/513/507 nm corresponding to the RTP of *L*-Lys/*L*-Ser/*L*-Ile are recorded, which remains rare even for aromatic pure organic luminogens.[18-23] Such RTP emission should be ascribed to the clustering−lowered energy gap ($\Delta E_{ST}$) between singlet excited states ($S_1$) and triplet excited states ($T_1$) together with enhanced spin-orbit coupling (SOC) and rigidified conformations. Preliminary theoretical calculation was also conducted using *L*-Ile as the model. From gas to crystals states, due to the involvement of π→σ* transition, the intersystem crossing is boosted with simultaneously impeded nonradiative deactivations from $T_1$ to ground state ($S_0$) (Figure S27, Supporting Information), thus generating remarkable RTP.

Upon cooling to 77 K, such solid emission is further enhanced, and moreover persistent phosphorescence is generated (Figure 3E), due to conformation rigidification and consequent highly suppressed molecular motions. Notably, for other native nonaromatic amino acids, wide emission spectra with fine and similar PL peaks covering from blue to red regions are also found (Figure S20–S23, Supporting Information), thus testifying the universal presence of heterogeneous emissive species. Specifically, with a $t_d$ of 0.1 ms, RTP emissions are detected for all compounds (Figure S24–S26, Supporting Information), indicating the prevalence of RTP emission in the crystalline solids. These results again reminder us that we should not ignore the contribution of aliphatic amino acids to the RTP emission of peptides and proteins, particularly at their solid states.

To further decipher the emission mechanism, single crystal structure which provides precise conformation and molecular packing of the compounds is analyzed. In the *L*-Ser crystals,[24] zwitterion structure is formed between carbonyl acid and amino moieties (**Figure 4**A). And the bond lengths of C1–O1 and C1–O2 are 1.262 and 1.250 Å, respectively, indicative of the electron delocalization over the whole carboxylate ion ($COO^-$). Furthermore, abundant intermolecular interactions including C–H···O=C, N–H···O=C, and O–H···O–H hydrogen bonds, C=O···C–H, O=C···H–N, H–O···O–H, and C=O···N–H short contacts are present around one molecule, forming a strong 3D intermolecular interaction network (Figure 4A and 4B). These intermolecular interactions, on one hand afford highly rigid molecular conformations, on the other hand, the O···O (2.907 Å) and C=O···N (2.820, 2.830, 2.861 Å) short contacts give rise to the through space electronic interactions (Figure 4A–4C). Exactly, an interconnected 3D through space electronic communication channel is formed in the *L*-Ser crystals (Figure 4D), which readily affords extended electron delocalization and subsequent visible light emission. It is noted that such 3D electronic communication channel is not solely restricted to specific compounds, but extensively existed in the crystals of nonaromatic amino acids. For example, there are two conformers in the *L*-Ile crystals,[25] in which zwitterion structure and electron delocalization of $COO^-$ groups are also found (Figure S28 and S29, Supporting Information). Other than abundant C=O···N electronic interactions (2.766, 2.775, 2.832, 2.846, 2.855, 2.962, 3.011 Å), there are also O=C···C=O ($\pi$-$\pi$, 3.332 Å) and O=C···O=C (n-$\pi^*$, 3.196 Å) communications (Figure S29, Supporting Information), which result in an effective 3D through space electronic communication channel and collectively contribute to generate optically excitable conjugates with rigidified conformations, thus offering bright visible emission upon UV illumination.

According to preceding results, it is rational to speculate that their corresponding polymers are also emissive. Normally, polymers possess more rigid conformations as the constraint of polymer chains, which also facilitate intra- and intermolecular interactions in concentrated solutions and solids. As a typical and commercially available poly(amino acid), $\varepsilon$-PLL was investigated. It is a water-soluble synthetic polypeptide which resembles basic proteins in some properties. In dilute solutions, $\varepsilon$-PLL chains assume random-coil conformations with isolated subgroups, which makes them virtually nonemissive [**Figure 5**A, $\Phi \approx 0$ ($10^{-3}$ mg mL$^{-1}$)], predominantly owing to the insufficient conjugation (Figure S30, Supporting Information). Their concentrated counterparts exhibit blue PL, whose intensity is boosted with increased concentration [Figure 5C, $\Phi = 13.8\%$ (15 mg mL$^{-1}$)]. PL peaks of the concentrated solutions also red-shift with much longer $\lambda_{ex}$ (Figure 5D). Meanwhile, $\varepsilon$-PLL powders demonstrate bright emission with $\Phi$ value of 7.9%, illustrating its AIE characteristics. Previously, people ascribed the emission of nonaromatic protein aggregates to the electron delocalization along the hydrogen bond.[5,6b] Considering the similar observations in diverse systems without typical hydrogen bonds,[8b] the origin of the emission can be ascribed to the clustering of nonconventional chromophores, which results in through space electronic communications among different π and n electrons. Hydrogen bonding, however, is beneficial to the conformation rigidification and facilitates the through space electronic communications of neighboring nonconventional chromophores, thus favoring for the emission.[8] With a $t_d$ of 0.1 ms, delay emissions corresponding to RTP are detected with maxima at 490~500 nm (Figure 5E). Strikingly, persistent RTP is clearly visualized after ceasing the irradiation (Figure 5B, Video S1, Supporting Information), which is scarcely found even for aromatic pure organic luminogens.[19-21] To the best of our knowledge, it is the first report of persistent RTP from nonconventional luminogens. Such unique RTP afterglow of $\varepsilon$-PLL indicates its highly rigid conformations, which should be ascribed to the polymer chain effect as well as strong intra- and intermolecular interactions. The occurrence of persistent RTP provides many opportunities for diverse applications like anti-counterfeiting and encryption.[20a,21]

Intrinsic emission and excellent biocompatibility of these nonaromatic amino acids and poly(amino acids) promoted us to explore their applications in bioimaging. As depicted in **Figure 6**, after incubation with 0.1 M *L*-Ile in Dulbecco's modified Eagle's medium (DMEM) for 1.5 h, HeLa cells exhibit bright blue emission under confocal microscopy with excitation at 405 nm, whereas no obvious fluorescent signal is detected for the control (Figure S31, Supporting Information). These results suggest *L*-Ile is ready to stain cells. Moreover, closer scrutinization reveals that *L*-Ile demonstrates specific imaging of endosomes, which is highly important to the biomedical research and clinical diagnosis.

In summary, unique intrinsic emission from nonaromatic amino acids and exampled poly(amino acid) of *ε*-PLL is observed. Whilst individual molecules are practically nonluminescent in dilute solutions even at 77 K, the majority of their concentrated counterparts and all solid powders demonstrate visible emission at room temperature even under UV-A irradiation. Clustering of nonconventional chromophores, which ensures subsequent electron cloud overlap, together with conformation rigidification is responsible for the emission. Single crystal analysis reveals the prevalence of such through space electronic communications as O···O, C=O···N, O=C···C=O, and O=C···O=C short contacts, which construct interpenetrated 3D networks. Besides the conformation rigidification effect, such 3D electronic communication channels are accountable for the extended delocalization and thus visible emission of the compounds. This CTE mechanism successfully correlates the previously discovered nonconventional luminogens (i.e. PAMAM, PAN, PU) with luminescent nonaromatic biomolecules. Furthermore, other than shortlived fluorescence, longlived and moreover persistent RTP are observed owing to the involvement of lone pairs, thus providing new implications for the RTP emission from proteins. We believe that the disclosure of the intrinsic emission of nonaromatic amino acids and poly(amino acids) paves the way for the design and elaboration of a new class of luminescent biomolecules with promising applications in biomedical and optoelectronic fields. Future investigation of these stuffs may help gain further insights into the kinetics of protein folding and the formation of

*β*-sheets at a label free manner, which are crucial to unveil the underlying mechanistic relations to disease and important to develop new diagnostic tools.

**Supporting Information**
Supporting Information is available from the Wiley Online Library or from the author.

**Acknowledgements**
This work was financially supported by the National Natural Science Foundation of China (51473092) and the Shanghai Rising-Star Program (15QA1402500). The authors appreciate Ms Xiaoli Bao and Ms Lingling Li at the Instrumental Analysis Center of Shanghai Jiao Tong University for the single-crystal structure determination of *L*-Ile.

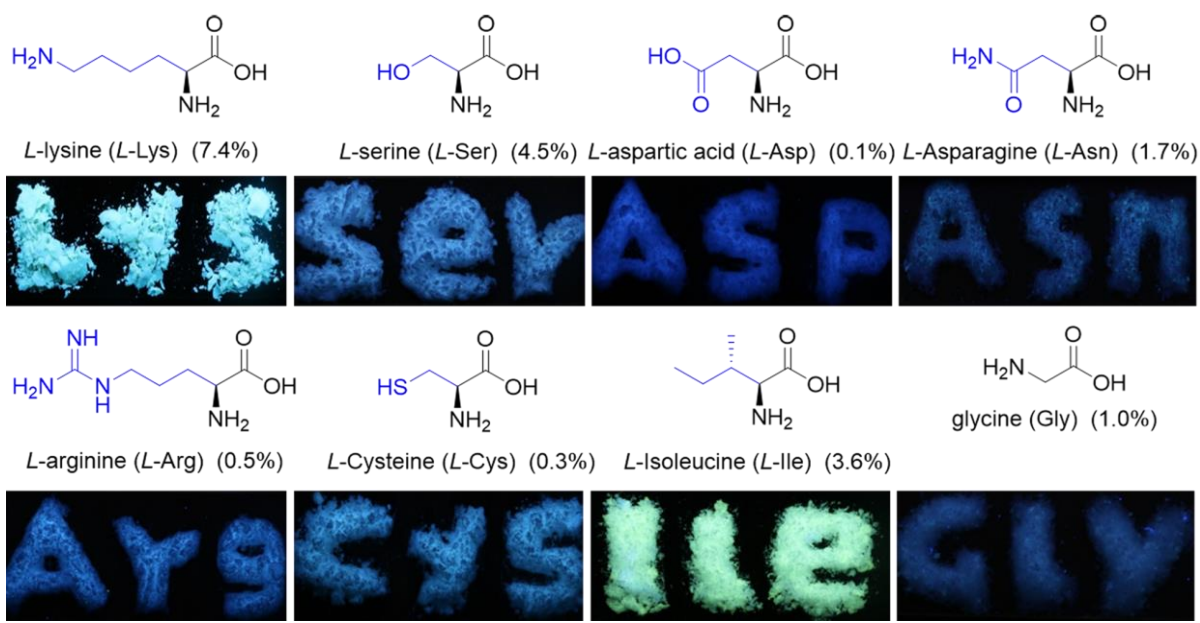

**Figure 1.** Exampled nonaromatic amino acids and photographs of the recrystallized solids taken under 365 nm UV light. Emission efficiencies of the recrystallized solids are given in the brackets.

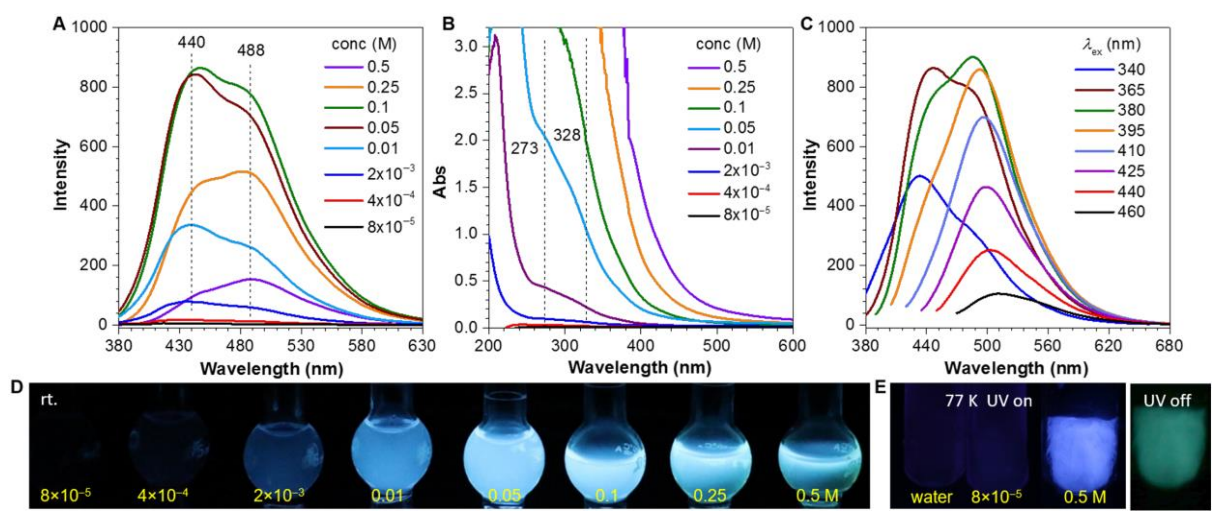

**Figure 2.** A) Emission ($\lambda_{ex}$ = 365 nm) and B) absorption spectra of varying *L*-Lys aqueous solutions. C) Emission spectra of 0.1 M *L*-Lys aqueous solution with different $\lambda_{ex}$s. Photographs of varying *L*-Lys aqueous solutions taken under 365 nm UV light or ceasing the irradiation at D) room temperature and E) 77 K.

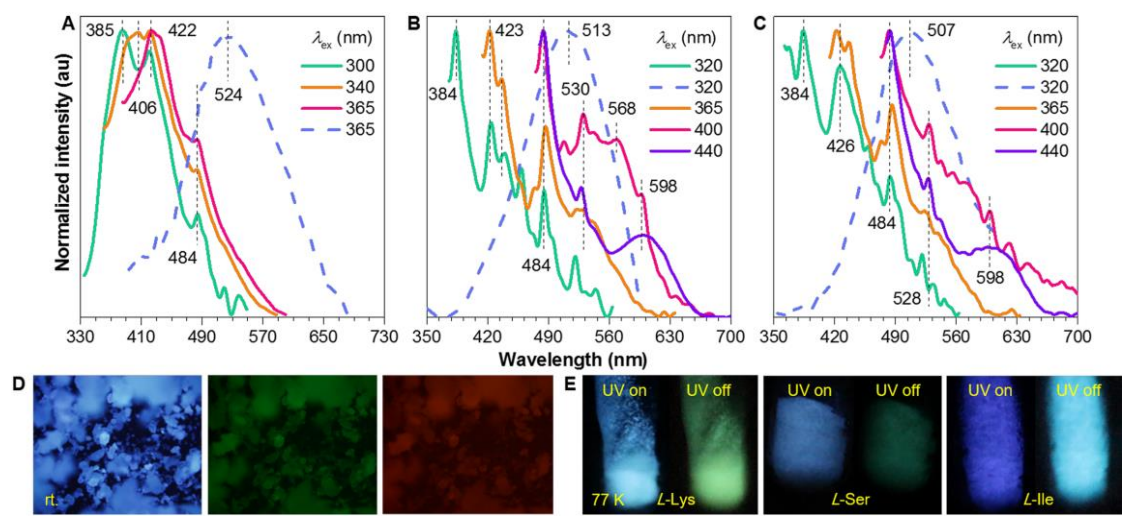

**Figure 3.** Emission spectra of recrystallized solids of A) *L*-Lys, B) *L*-Ser, and C) *L*-Ile with $t_d$ of 0 (solid line) and 0.1 ms (dash line). D) Microscope images of *L*-Lys solids taken under illumination of UV (330–385 nm, left), blue (460–495 nm, middle) and green (530–550 nm, right) lights. E) Photographs of *L*-Lys, *L*-Ser, and *L*-Ile solids taken at 77 K under 365 nm UV light or after ceasing the UV irradiation.

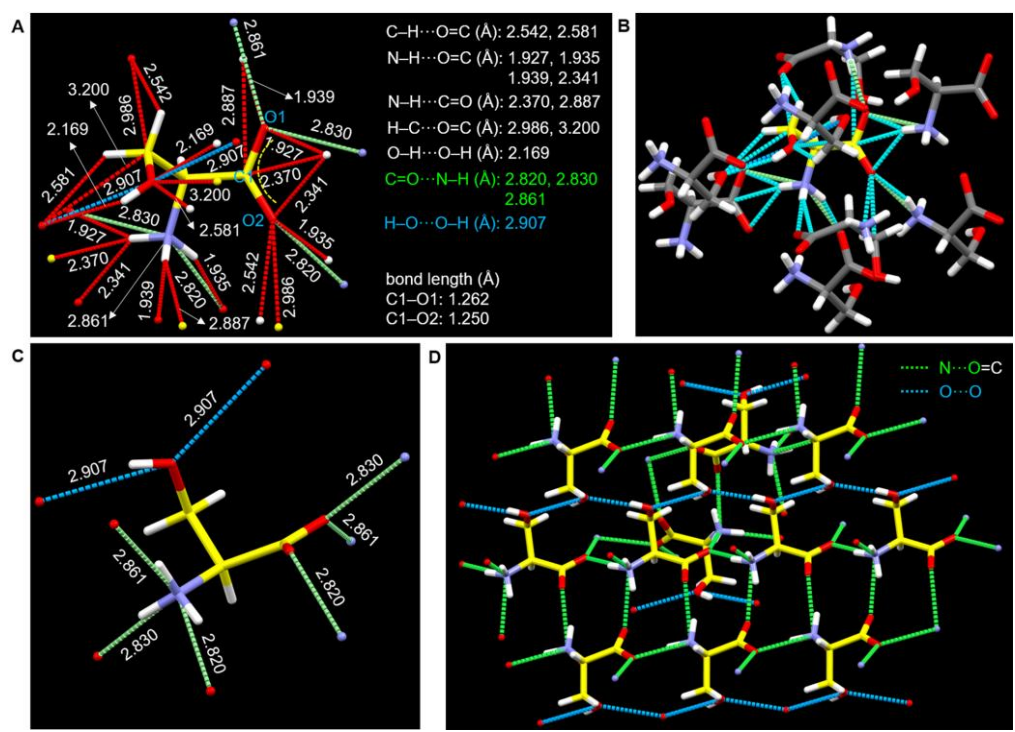

**Figure 4.** A, B) Crystal structure of *L*-Ser with denoted intermolecular interactions around one molecule. C) N⋯O and O⋯O intermolecular interactions around one molecule. D) Fragmental 3D through space electronic communication channel in the *L*-Ser crystals.

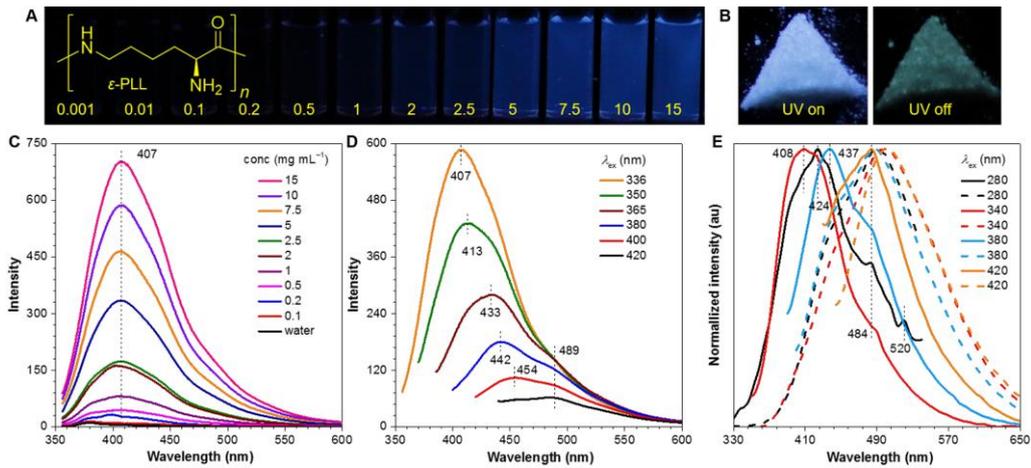

**Figure 5.** Photographs of A) different ε-PLL aqueous solutions and B) solid powders taken under 365 nm UV light or after ceasing the UV irradiation. Emission spectra of C) different ε-PLL aqueous solutions ($\lambda_{ex}$ = 336 nm) and D) 15 mg mL$^{-1}$ solution with varying $\lambda_{ex}$s. E) Normalized emission spectra of ε-PLL solids with $t_d$ of 0 (solid line) and 0.1 ms (dash line) under varying $\lambda_{ex}$s.

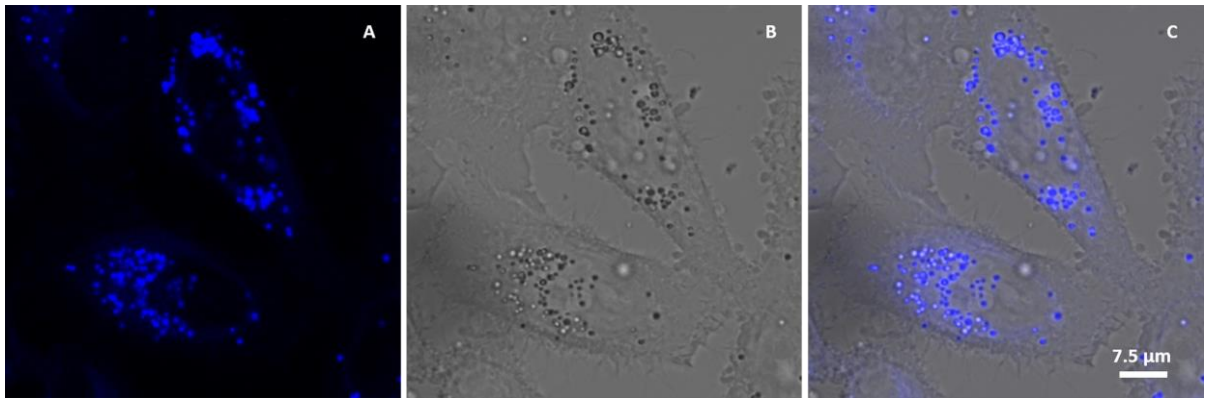

**Figure 6.** Confocal luminescent images of HeLa cells after incubation with 0.1 M *L*-Ile in DMEM for 1.5 h. A) Confocal image recorded under excitation at 405 nm, B) bright field image, and C) corresponding overlayed image.

# Supporting Information

**Prevalent Intrinsic Emission from Nonaromatic Amino Acids and Poly(Amino Acids)**

*Xiaohong Chen, Weijian Luo, Huili Ma, Qian Penguin, Wang Zhang Yuan\*, and Yongming Zhang\**

**Experimental Section**

  *Materials*: Amino acids were purchased from J&K Scientific Chemical Co., Ltd. (Shanghai, China). *ε*-Poly-*L*-lysine (*ε*-PLL, $M_w$ < 5000) was obtained from Macklin Biochemical Co., Ltd. (Shanghai, China). Deionized water was used throughout the recrystallization of amino acids and *ε*-PLL. The pure water was bought from Hangzhou Wahaha Group Co., Ltd. (Zhejiang, China), and was used for the preparation of aqueous solutions of nonaromatic amino acids and *ε*-PLL. Methanol and ethanol were provided by Yonghua Chemical Technology Co., Ltd. (Jiangsu, China) and Changshu Hongsheng Fine Chemical Co., Ltd. (Jiangsu, China), respectively. Barium sulfate ($BaSO_4$, AR) and anhydrous diethyl ether were obtained from Sinopharm Chemical reagent Co., Ltd. (Shanghai, China). Dulbecco's modified Eagle's medium (DMEM) was obtained from Thermo Scientific (USA).

  *Recrystallization of Amino Acids*: (1) Gly, *L*-Ala, *D*-Ala, *L*-Ser, *D*-Ser, *L*-Arg, *D*-Arg, and *L*-Lys: These amino acids were firstly dissolved in deionized water at room temperature, then bad solvent (methanol or ethanol) was added to yield the precipitates. (2) *L*-Pro: It was firstly dissolved in the mixture of deionized water and ethanol. Upon addition of diethyl ether anhydrous, it precipitated out from the mixture. (3) The other amino acids: Due to the relatively low solubility in water, these amino acids were firstly dissolved in water at 90 °C, then bad solvent (methanol or ethanol) was added. Solids came out from the mixtures with further cooling with an ice-water bath. All above precipitates were filtrated with sand-core funnels. After being dried in a vacuum oven at 40 °C overnight, they were used for further experiments.

*Purification of ε-PLL*: ε-PLL was firstly dissolved in water (W). Then it was purified by precipitation in ethanol (E) (W/E, 1:10 by volume). After centrifugation, the collected solid was dried in vacuum at 40 °C overnight for further characterization.

*Preparation of L-Ile-L-Arg and L-Ile-Gly Solid Mixtures*: Recrystallized solids of *L*-Ile and *L*-Arg (or Gly) with the molar ratio of 1:1 were dissolved in water at 90 °C. After filtration, the mixture solution was collected into a watch glass. Then it was dried in a vacuum oven at 40 °C overnight.

*Single Crystal Cultivation of L-Ile*: *L*-Ile crystal was obtained by solvent evaporation in pure water solution. Into a sample bottle (5 mL) were added of *L*-Ile powders (5 mg) and pure water (2 mL). The bottle was sealed by a rubber plug with a pinhead. Resulting *L*-Ile crystal was obtained after standing for one month.

*Characterization Methods*: Absorption spectra of solutions and solids of all amino acids and ε-PLL were taken on a Lambda 35 UV/Vis spectrometer (Perkin Elmer, USA) spectrometer and UV-2450 UV-Vis spectrophotometer (Shimadzu, Japan), respectively. Excitation and emission spectra were determined at room temperature on a Perkin-Elmer LS 55 fluorescence spectrometer (PerkinElmer, USA). The lifetimes of fluorescence were acquired with a QM/TM/IM steady-transient time-resolved spectroscopy (PTI, USA). A PL-F2X nitrogen laser as the excitation source for lifetime measurements, with the resulting fluorescence monitored at right angles through an emission monochromtor with a photomultiplier tube and strobe detector. Quantum yields of the solutions and solids of *L*-Lys, *L*-Ser, *L*-Ile, and ε-PLL were measured on a spectrophotometer (PTI, USA) equipped with SPEKTRON-R98 coated integrating sphere ($\varphi$ 80 mm) (Everfine, China), with $\lambda_{ex}$ of 336 (ε-PLL solutions), 340 (*L*-Lys, *L*-Ser, *L*-Ile solutions and solids), and 350 (ε-PLL solids) nm. Quantum efficiencies of the recrystallized solids of the other amino acids were determined on a Quantaurus-QY C11347-11 absolute PL quantum yield measurement system (Hamamatsu, Japan), with $\lambda_{ex}$ of 330 (Gly) and 365 nm (others). All photographs were taken by a digital camera (Canon EOS 70D, Japan). The mean sizes of the aggregates of *L*-Lys, *L*-Ser and *L*-Ile in water were determined at 25 °C by DLS using a Malvern ZS90 (ZEN 3690) instrument (Malvern, UK). Photographs of recrystallized solids of *L*-Lys excited by different lights of 330–385 (UV), 460–495 (blue) and 530–550 nm (green) were recorded on a Reflected Fluorescence System (Olympus BX61, Japan). Crystallography data for *L*-Ile were collected on a Bruker D8 Venture-CMOS diffractometer with Cu Kα radiation ($\lambda$=1.54178 Å) at 173 K. The structure was solved by direct methods and refined by the full-matrix least-squares method on $F^2$ using the SHELXTL 2014 crystallographic software package. Anisotropic thermal parameters were used to refine all non-hydrogen atoms. All hydrogen atoms were placed in the riding model and refined isotropically.

*Cell Culture and Confocal Imaging*: HeLa cells were seeded in a 4-chamber glass bottom dish (35 mm dish with 20 mm bottom well) with DMEM (high glucose, 1% *L*-glutamine) containing 10% fetal bovine serum and 1% penicillin-streptomycin. The dish was cultured in a humidified incubator containing 5% $CO_2$ at 37 °C for 24 h, then the cells were stained with *L*-Ile/DMEM solution (500 μL, 0.1 M) for 1.5 h. Afterwards, they were imaged using a laser scanning confocal microscopy (Leica SP8 STED 3X) at an excitation wavelength of 405 nm.

*Computational Details*: The computational models were built from the crystal structure shown in **Chart S1**. The quantum mechanics/molecular mechanics (QM/MM) theory with two-layer ONIOM method was implemented to deal with the electronic structures in crystal, where the central molecule is chosen as the active QM part and set as the high layer, while the surrounding ones are chosen as the MM part and defined as the low layer. The universal force field (UFF) was used for the MM part, and the molecules of MM part were frozen during the QM/MM geometry optimizations. On the basis of the optimized geometry of the ground state ($S_0$) both in gas and solid at B3LYP/6-311G(d,p) level, the excitation energies were calculated by using TD-DFT for electronic excited singlet and triplet states. The above results are calculated by Gaussian 09 package.[S1] At the same level, the spin-orbit coupling between singlet and triplet states, as well as the oscillator strength of triplet states are given by Beijing Density Function (BDF) program.[S2-S4]

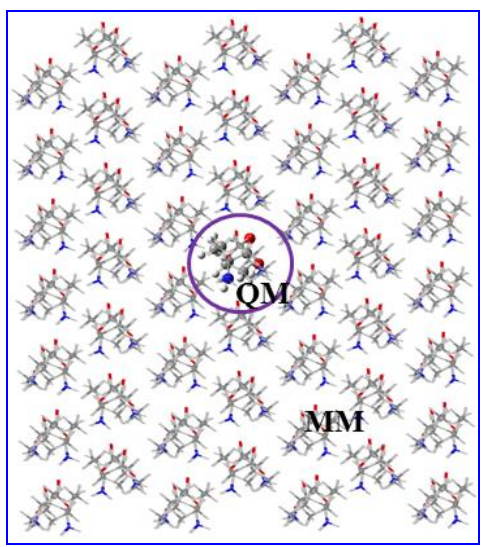

**Chart S1.** QM/MM model taking *L*-Ile as an example: one central QM molecule for the higher layer and the surrounding 293 MM molecules for the lower layer.

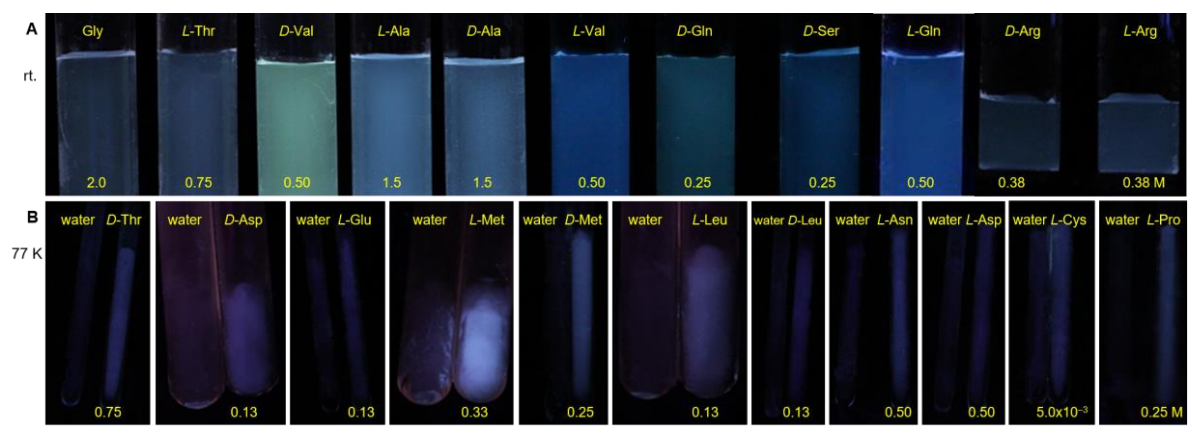

**Figure S1.** Photographs of different concentrated nonaromatic amino acids taken under 365 nm UV light at A) room temperature (rt) or B) 77 K. For the nonaromatic amino acids in B), no visible emission is observed at room temperature.

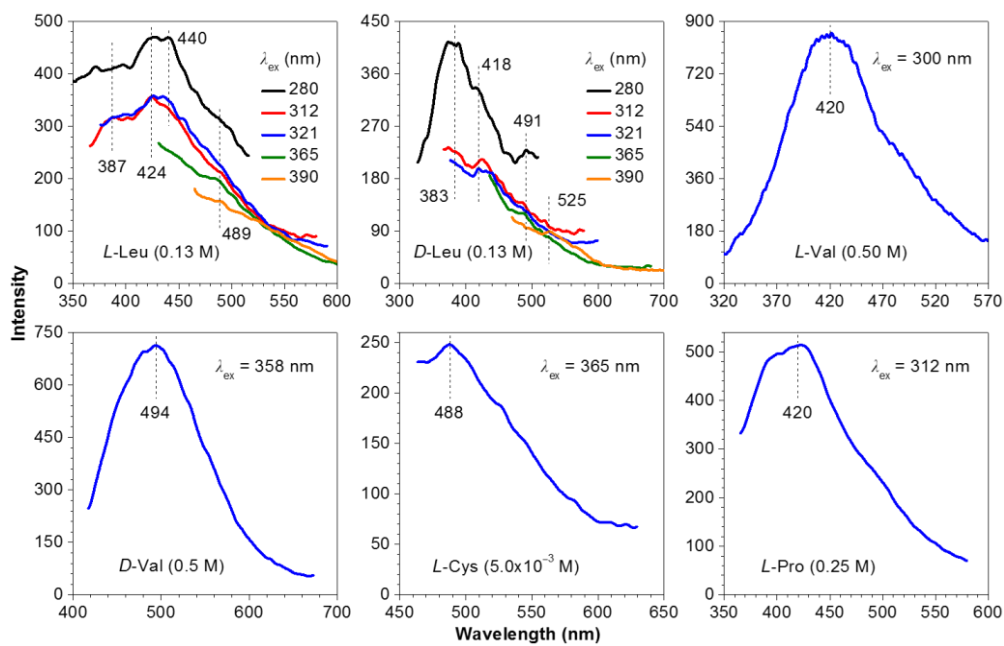

**Figure S2.** Emission spectra of *L*-Leu, *D*-Leu, *L*-Val, *D*-Val, *L*-Cys, and *L*-Pro in water with concentration and excitation wavelength ($\lambda_{ex}$) indicated.

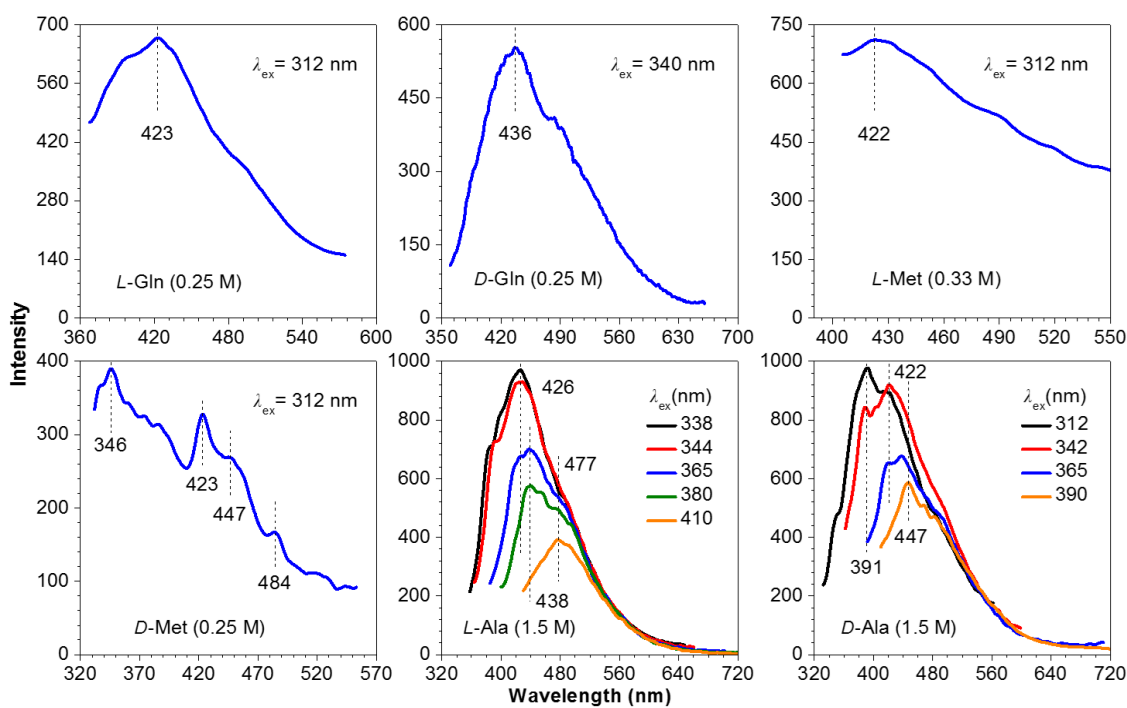

**Figure S3.** Emission spectra of *L*-Gln, *D*-Gln, *L*-Met, *D*-Met, *L*-Ala, and *D*-Ala in water with concentration and $\lambda_{ex}$ indicated.

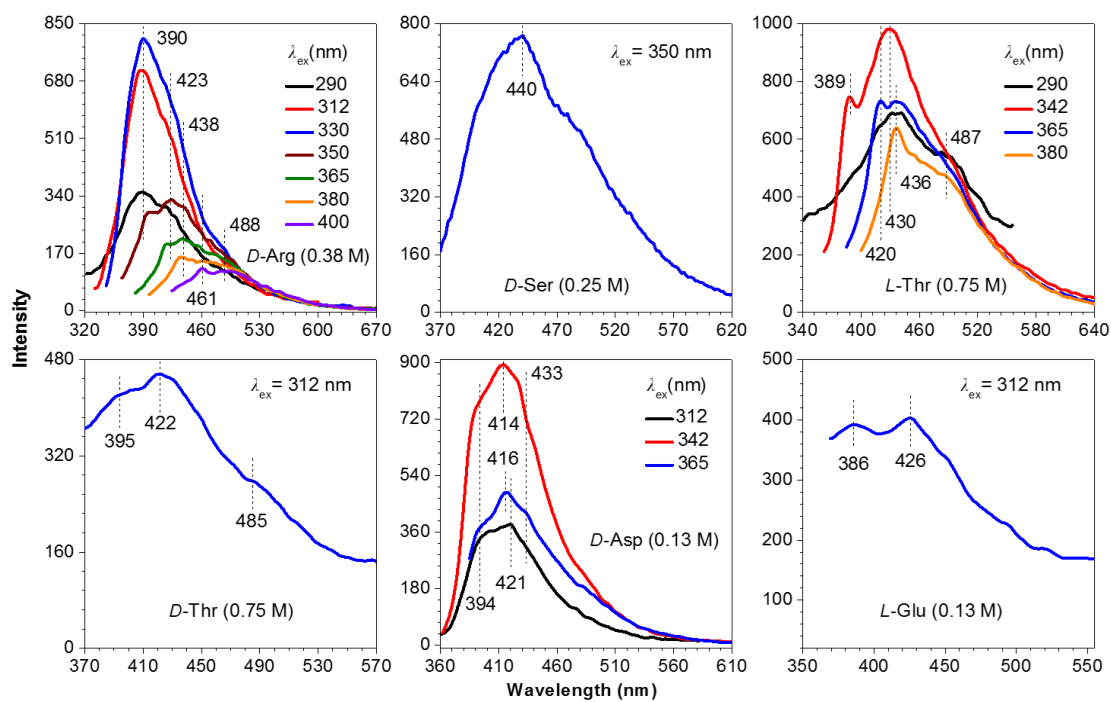

**Figure S4.** Emission spectra of *D*-Arg, *D*-Ser, *L*-Thr, *D*-Thr, *D*-Asp, and *L*-Glu in water with concentration and $\lambda_{ex}$ indicated.

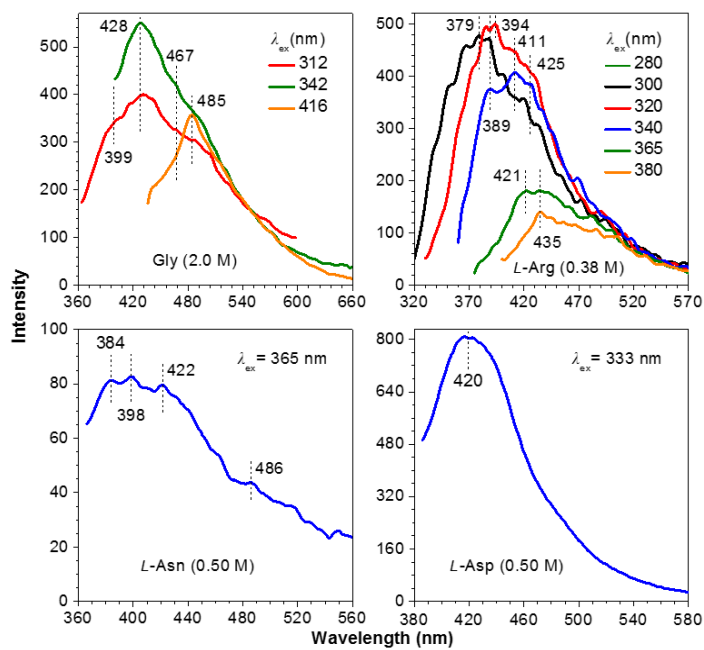

**Figure S5.** Emission spectra of Gly, *L*-Arg, *L*-Asn, and *L*-Asp in water with concentration and $\lambda_{ex}$ indicated.

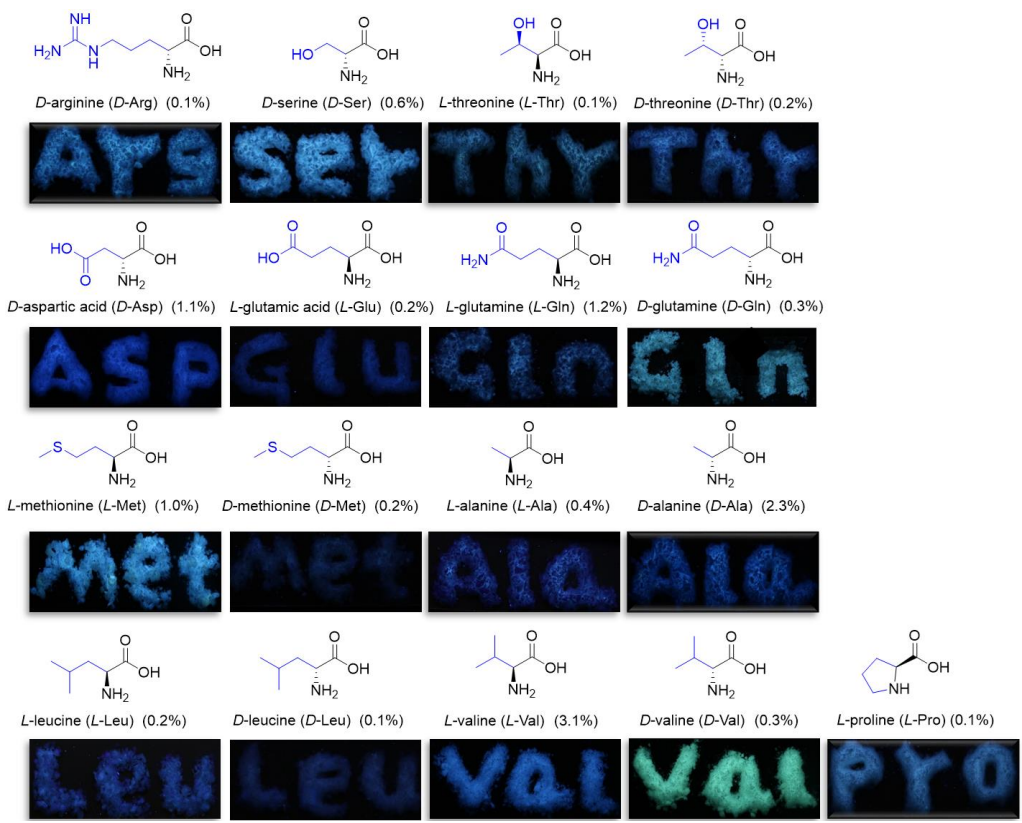

**Figure S6.** Chemical structures of some nonaromatic amino acids and photographs of their recrystallized solids taken under 365 nm UV light. Emission efficiencies of the recrystallized solids are given in the brackets.

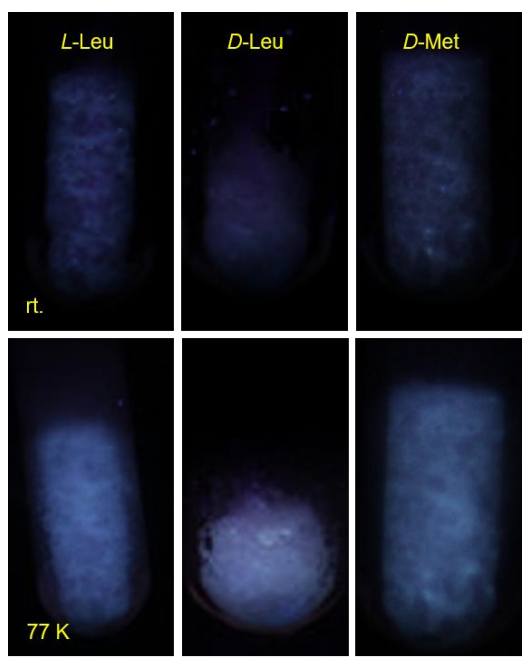

**Figure S7.** Photographs of *L*-Leu, *D*-Leu, and *D*-Met taken under 365 nm UV light at room temperature (upper) and 77 K (lower).

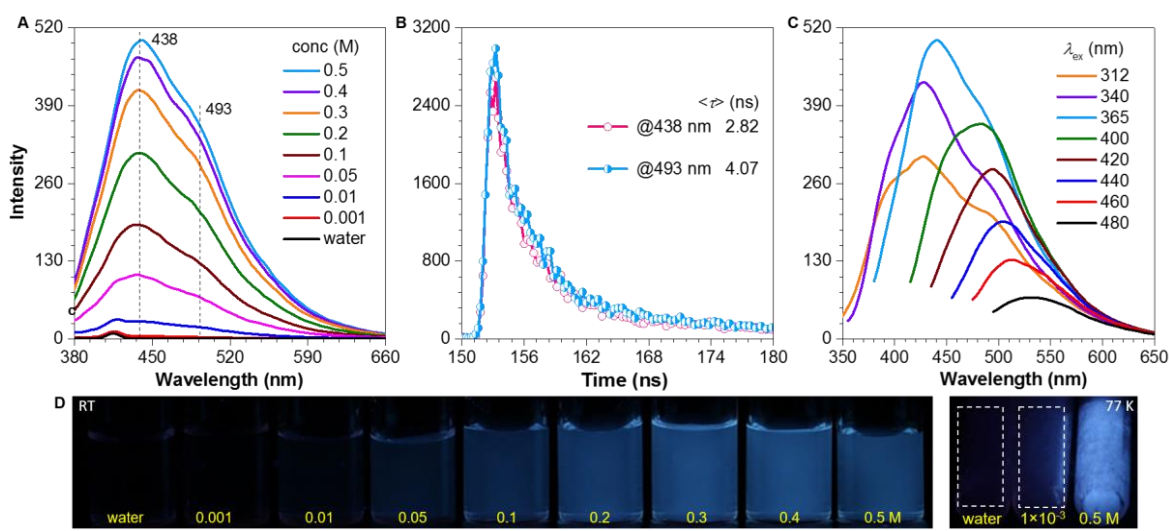

**Figure S8.** A) Emission spectra of varying *L*-Ser aqueous solutions ($\lambda_{ex}$ = 365 nm). B) Lifetimes of 0.5 M *L*-Ser solution monitored at 438 and 493 nm ($\lambda_{ex}$ = 365 nm). C) Emission spectra of 0.5 M *L*-Ser solution with different $\lambda_{ex}$s. D) Photographs of water and varying *L*-Ser aqueous solutions taken under 365 nm UV light at room temperature (left) and 77 K (right).

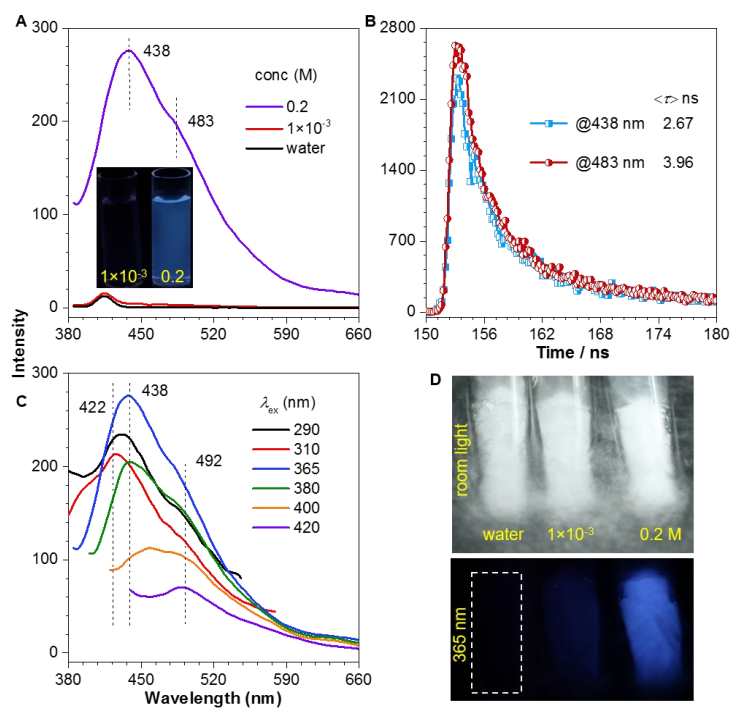

**Figure S9.** A) Emission spectra of varying *L*-Ile aqueous solutions ($\lambda_{ex}$ = 365 nm). B) Lifetimes of 0.2 M *L*-Ile solution monitored at 438 and 483 nm ($\lambda_{ex}$ = 365 nm). C) Emission spectra of 0.2 M *L*-Ser aqueous solution with different $\lambda_{ex}$s. D) Photographs of water and *L*-Ile aqueous solutions taken at 77 K under room light (top) and 365 nm UV light (bottom).

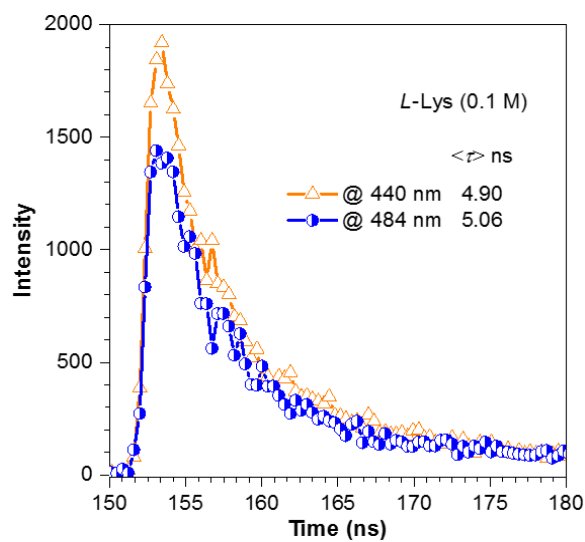

**Figure S10.** Lifetimes of *L*-Lys aqueous solution (0.1 M) monitored at 440 and 484 nm ($\lambda_{ex}$ = 365 nm).

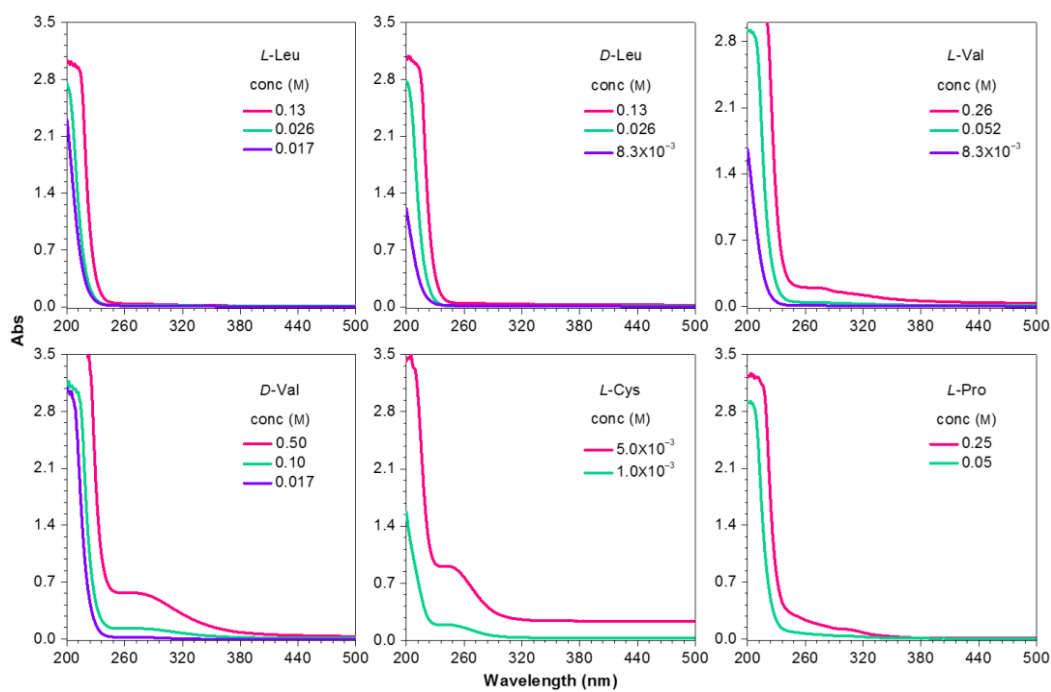

**Figure S11.** Absorption spectra of *L*-Leu, *D*-Leu, *L*-Val, *D*-Val, *L*-Cys, and *L*-Pro in water at different concentrations.

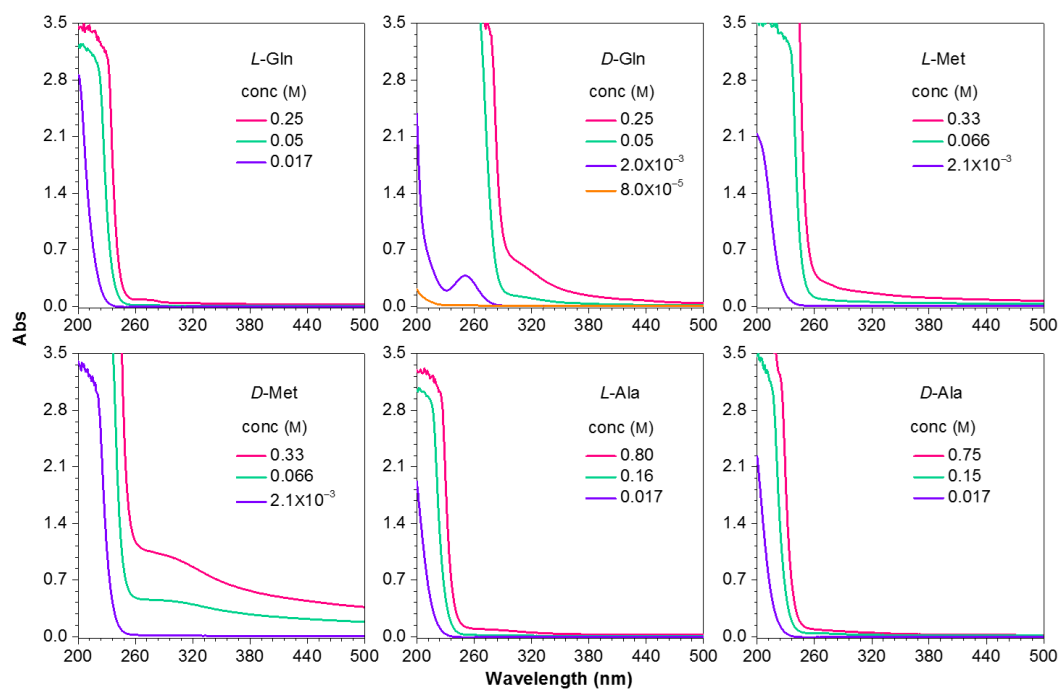

**Figure S12.** Absorption spectra of *L*-Gln, *D*-Gln, *L*-Met, *D*-Met, *L*-Ala, and *D*-Ala in water at different concentrations.

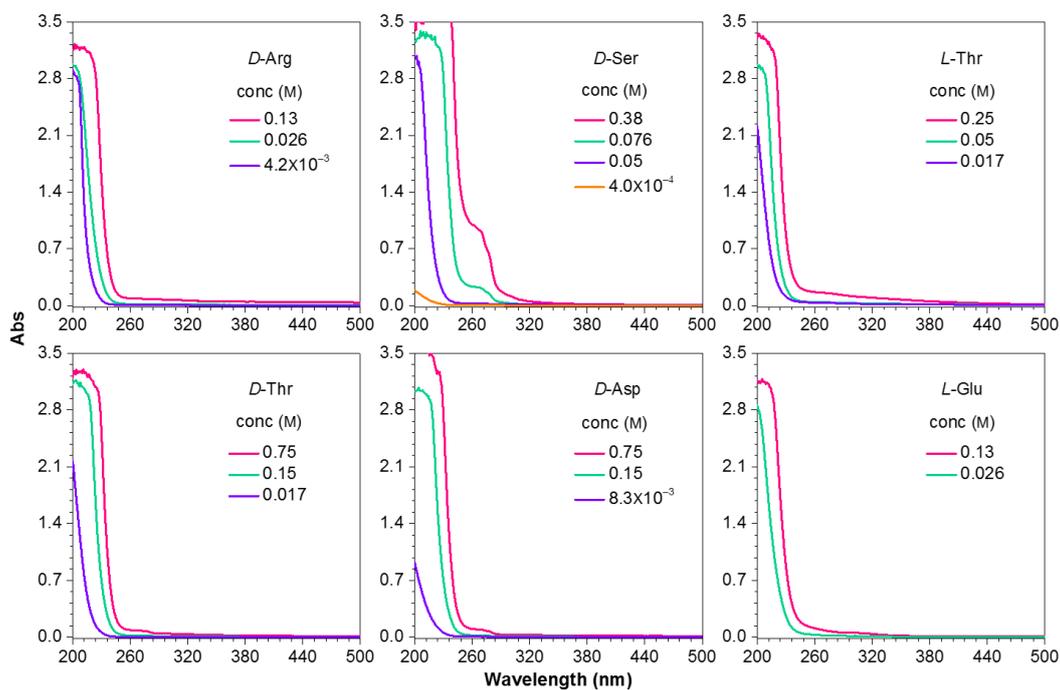

**Figure S13.** Absorption spectra of *D*-Arg, *D*-Ser, *L*-Thr, *D*-Thr, *D*-Asp, and *L*-Glu in water at different concentrations.

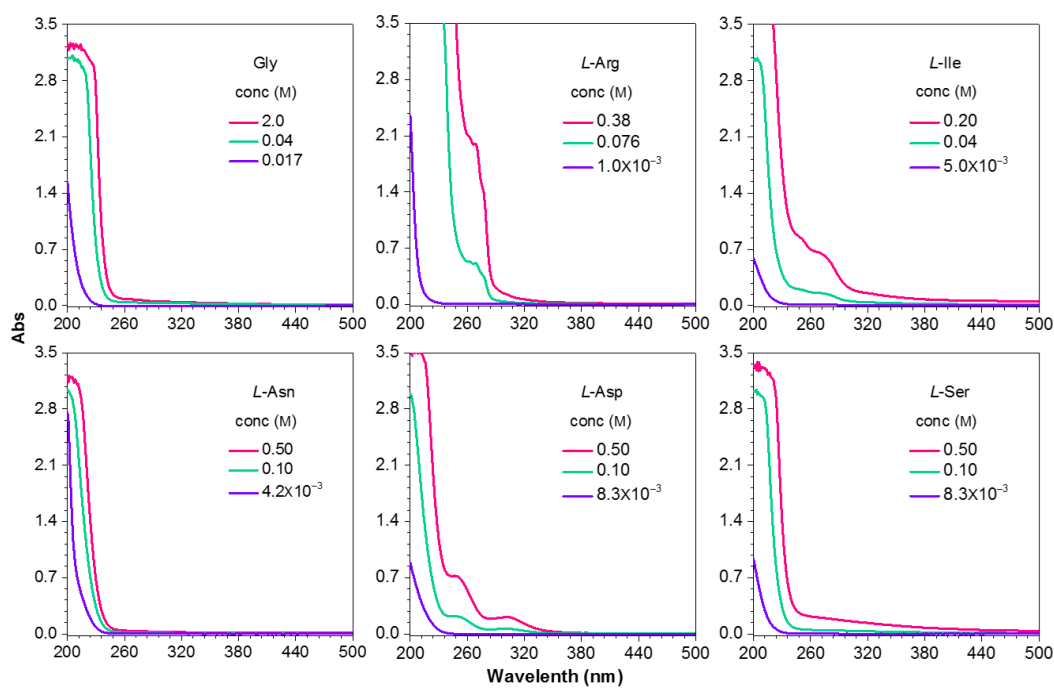

**Figure S14.** Absorption spectra of Gly, *L*-Arg, *L*-Ile, *L*-Asn, *L*-Asp, and *L*-Ser in water at different concentrations.

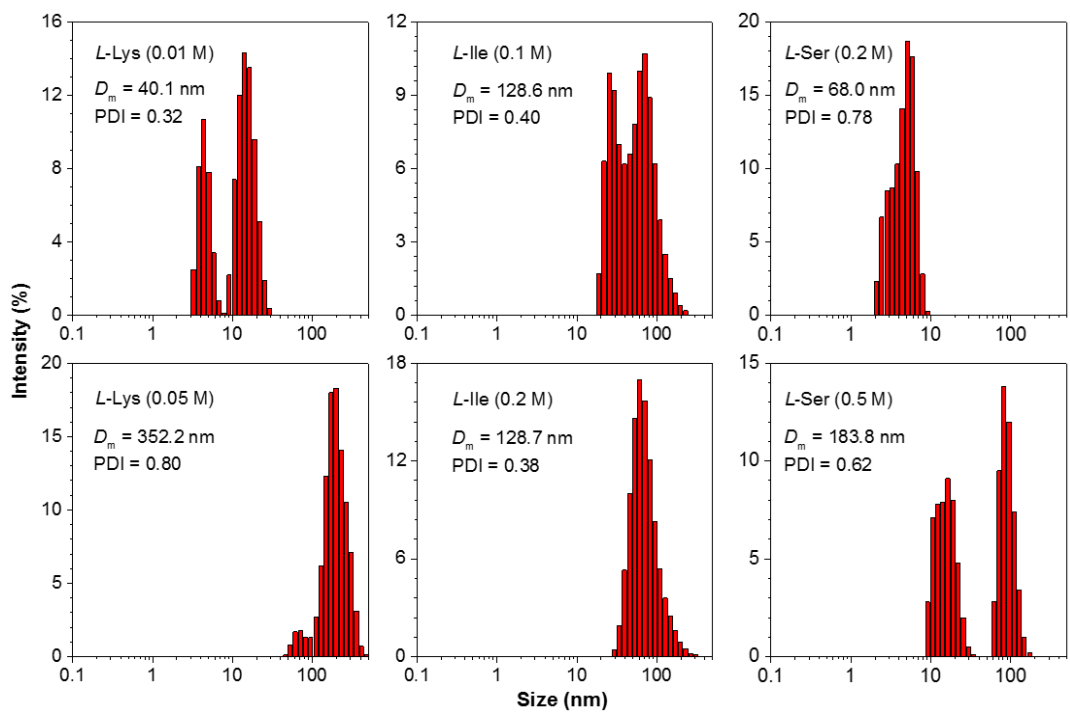

**Figure S15.** DLS data of *L*-Lys, *L*-Ile, and *L*-Ser aqueous solutions at different concentrations as indicated. $D_m$ = mean diameter; PDI = polydispersity index.

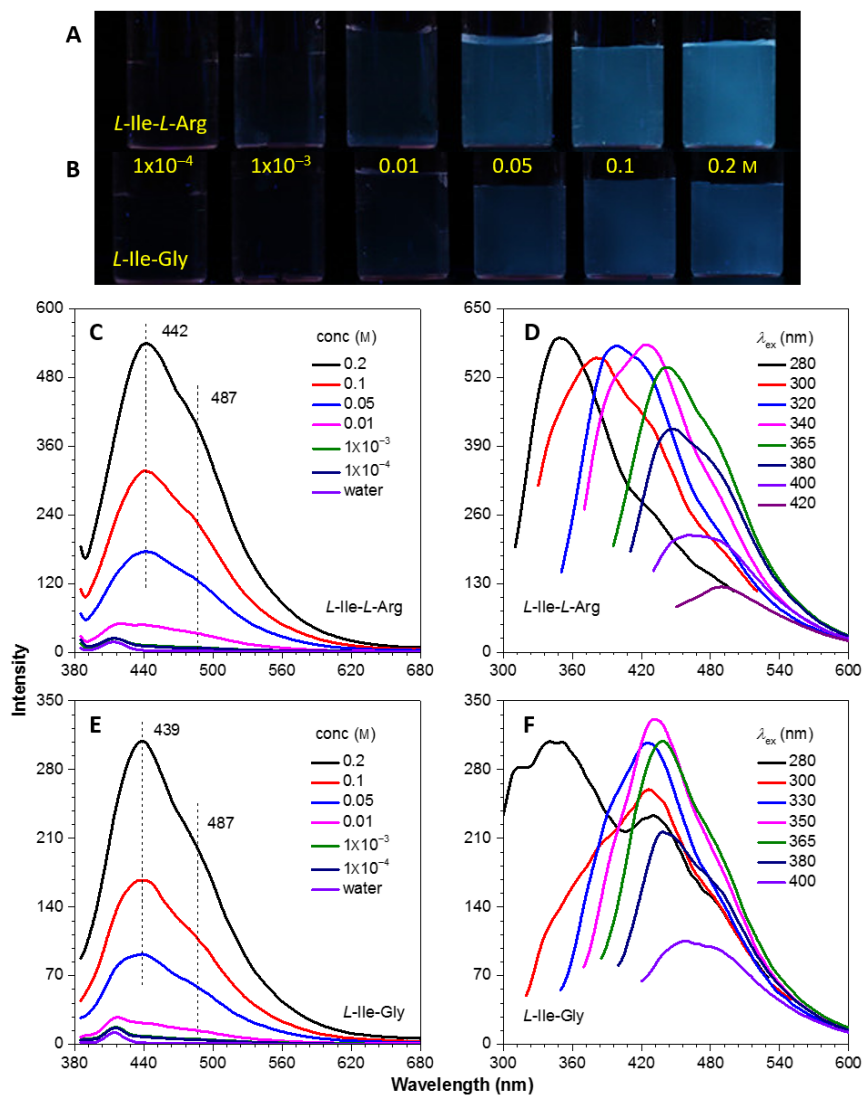

**Figure S16.** A, B) Photographs taken under 365 nm UV light and C–F) emission spectra of different aqueous solutions of A, C, D) *L*-Ile-*L*-Arg and B, E, F) *L*-Ile-Gly. [*L*-Ile]: [*L*-Arg] = [*L*-Ile]: [Gly] = 1:1. The indicated concentration is that for each single amino acid in the solution. Sample concentration in D) and F) is 0.2 M.

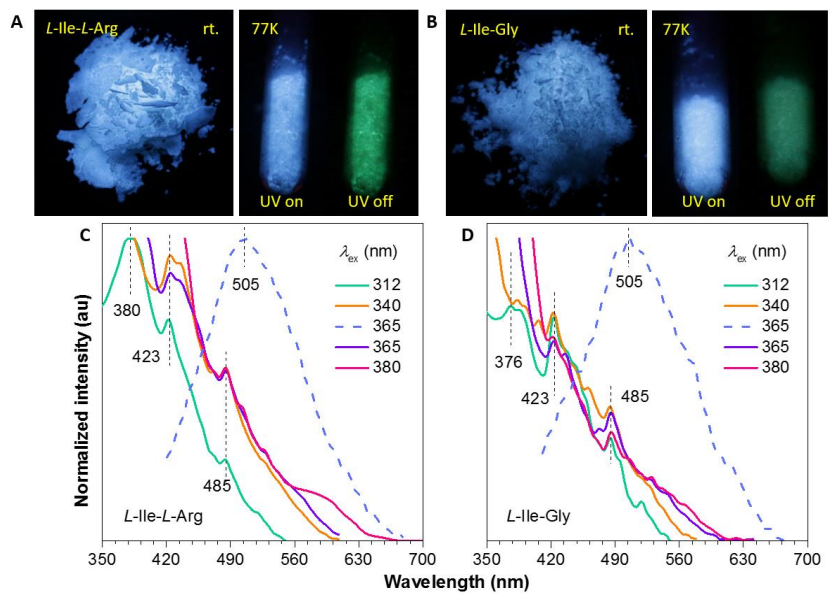

**Figure S17.** A, B) Photographs for the solid powders of A) *L*-Ile-*L*-Arg and B) *L*-Ile-Gly taken under 365 nm UV light or after ceasing the UV irradiation at room temperature and 77 K. C, D) Emission spectra for the solid powders of C) *L*-Ile-*L*-Arg and *L*-Ile-Gly with delay time of 0 (solid line) and 0.1 ms (dash line) under varying $\lambda_{ex}$s. [*L*-Ile]: [*L*-Arg] = [*L*-Ile]: [Gly] = 1:1.

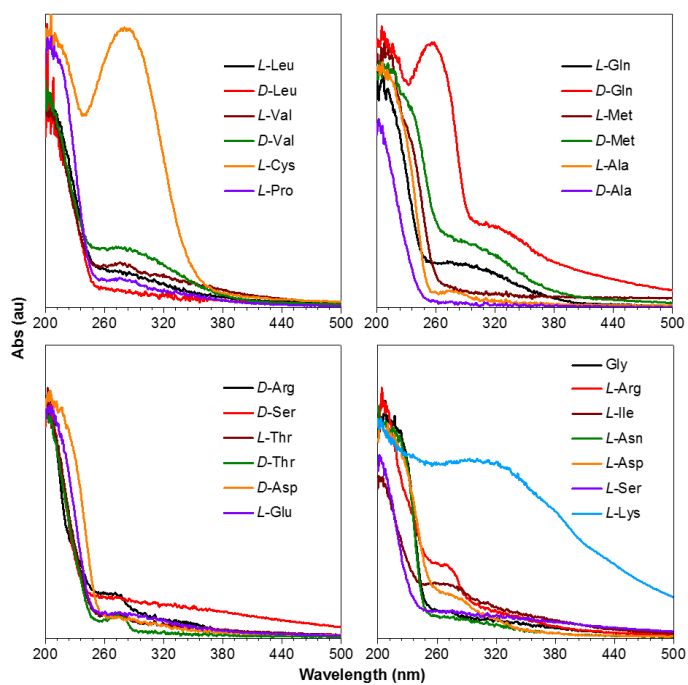

**Figure S18.** Absorption spectra of the recrystallized solids of different nonaromatic amino acids.

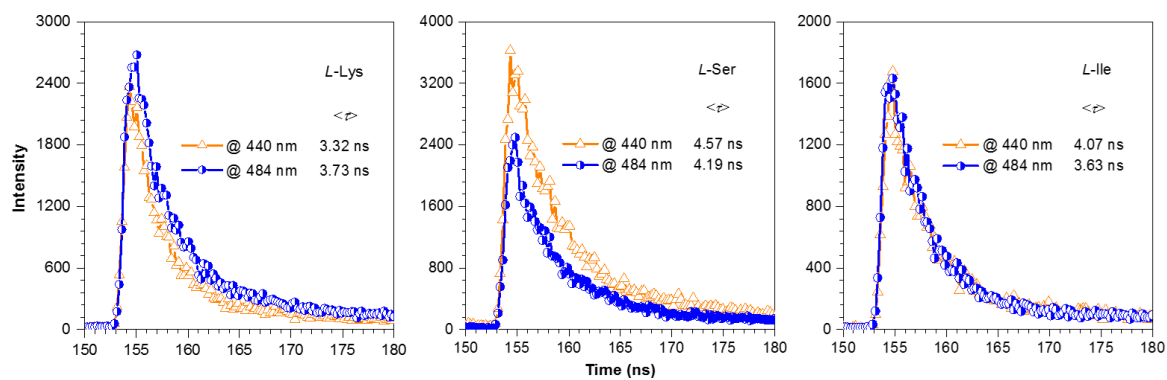

**Figure S19.** Lifetimes of *L*-Lys, *L*-Ser and *L*-Ile powders under the excitation wavelength at 365 nm.

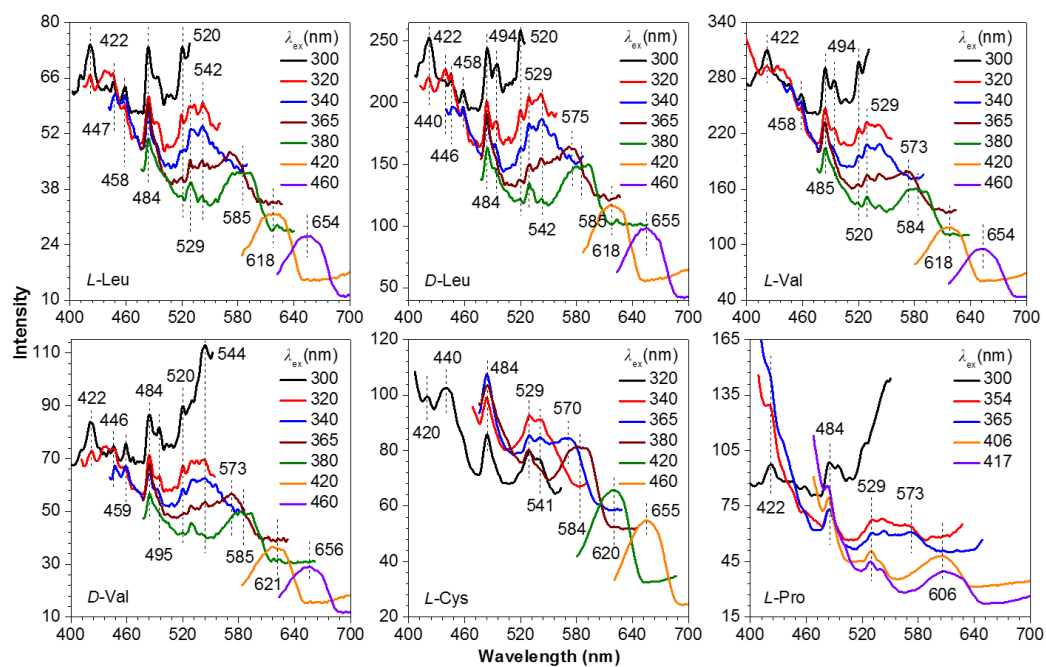

**Figure S20.** Emission spectra of recrystallized solids of *L*-Leu, *D*-Leu, *L*-Val, *D*-Val, *L*-Cys, and *L*-Pro with different $\lambda_{ex}$s.

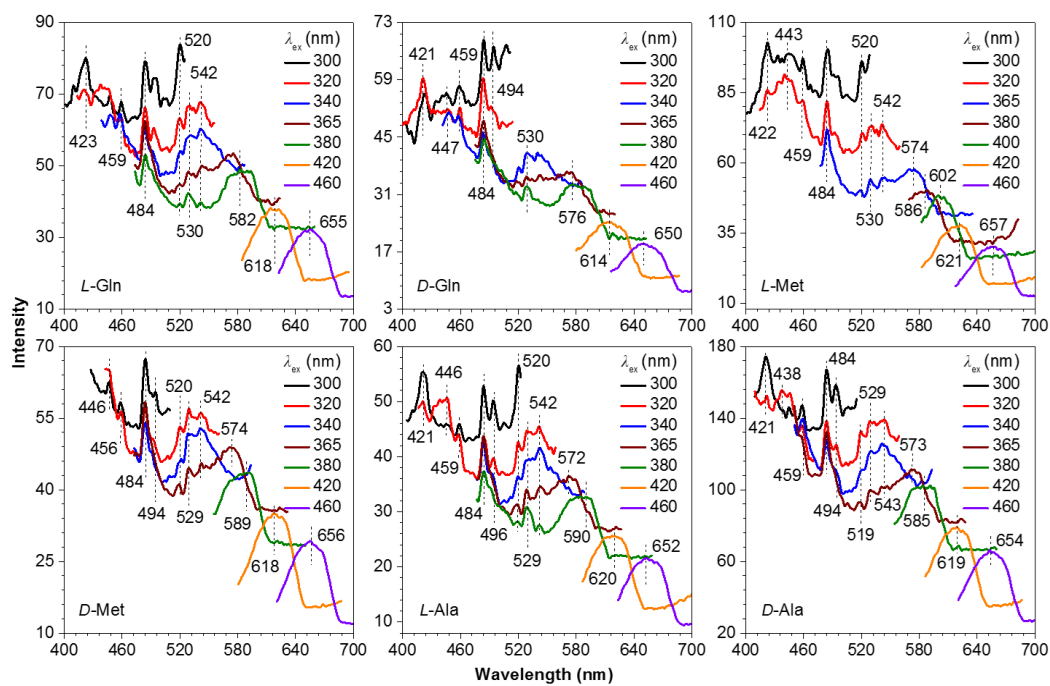

**Figure S21.** Emission spectra of recrystallized solids of *L*-Gln, *D*-Gln, *L*-Met, *D*-Met, *L*-Ala, and *D*-Ala with different $\lambda_{ex}$s.

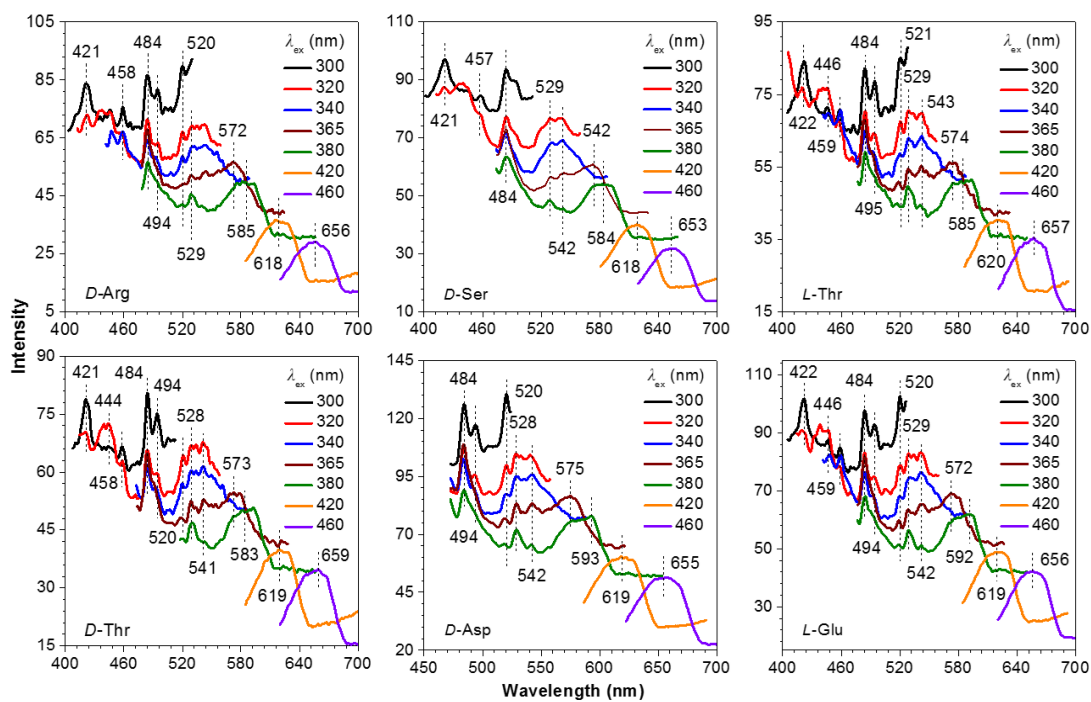

**Figure S22.** Emission spectra of recrystallized solids of *D*-Arg, *D*-Ser, *L*-Thr, *D*-Thr, *D*-Asp, and *L*-Glu with different $\lambda_{ex}$s.

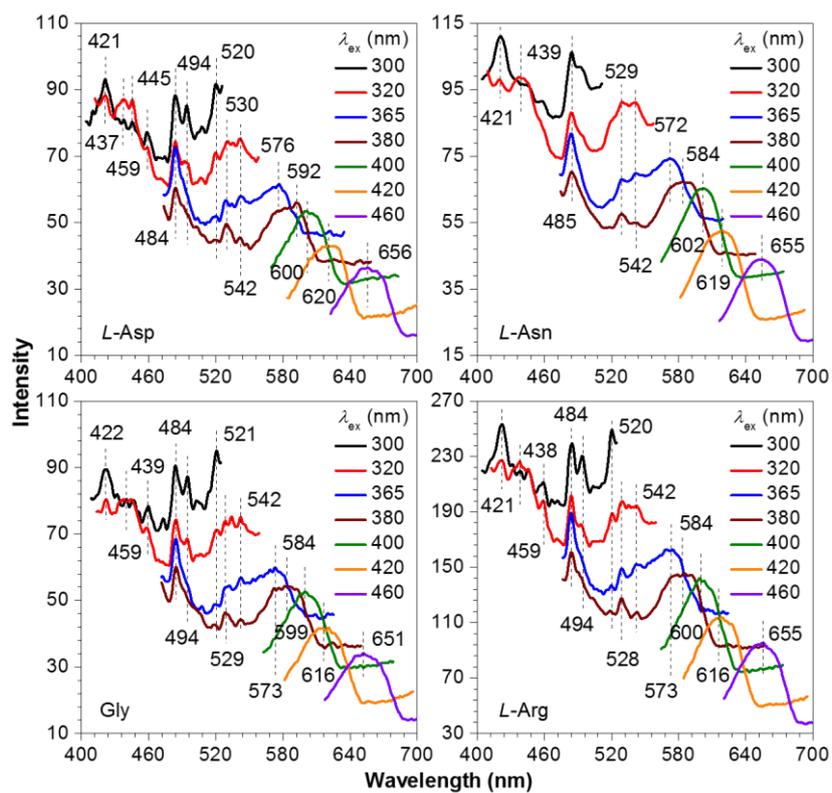

**Figure S23.** Emission spectra of recrystallized solids of *L*-Asp, *L*-Asn, Gly, and *L*-Arg with different $\lambda_{ex}$s.

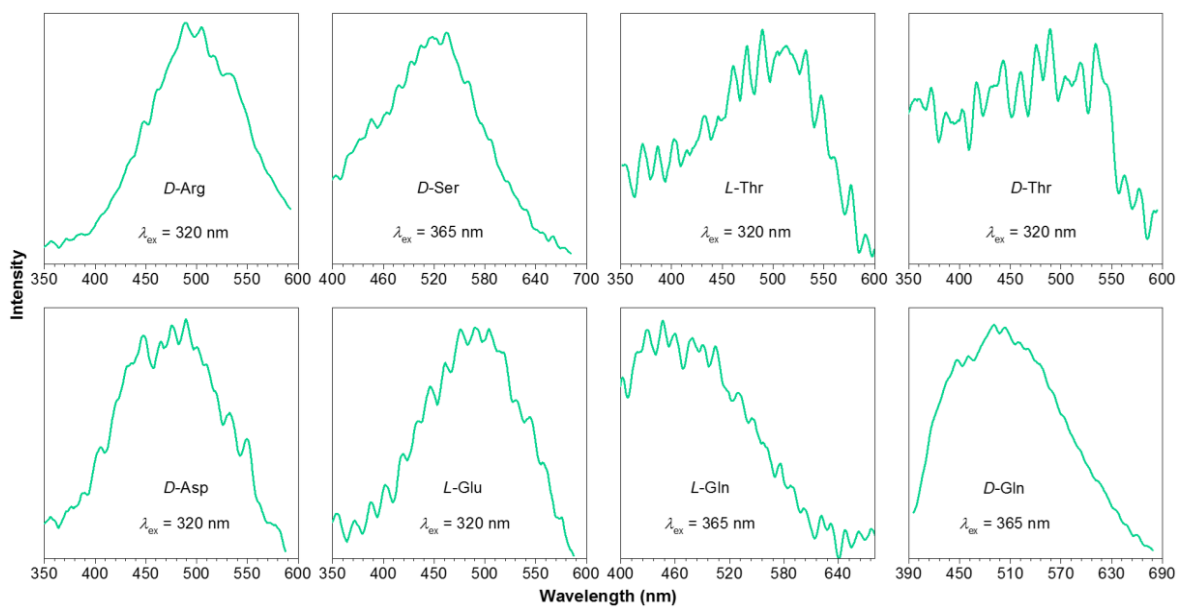

**Figure S24.** Emission spectra of recrystallized solids of *D*-Arg, *D*-Ser, *L*-Thr, *D*-Thr, *D*-Asp, *L*-Glu, *L*-Gln, and *D*-Gln with delay time ($t_d$) of 0.1 ms.

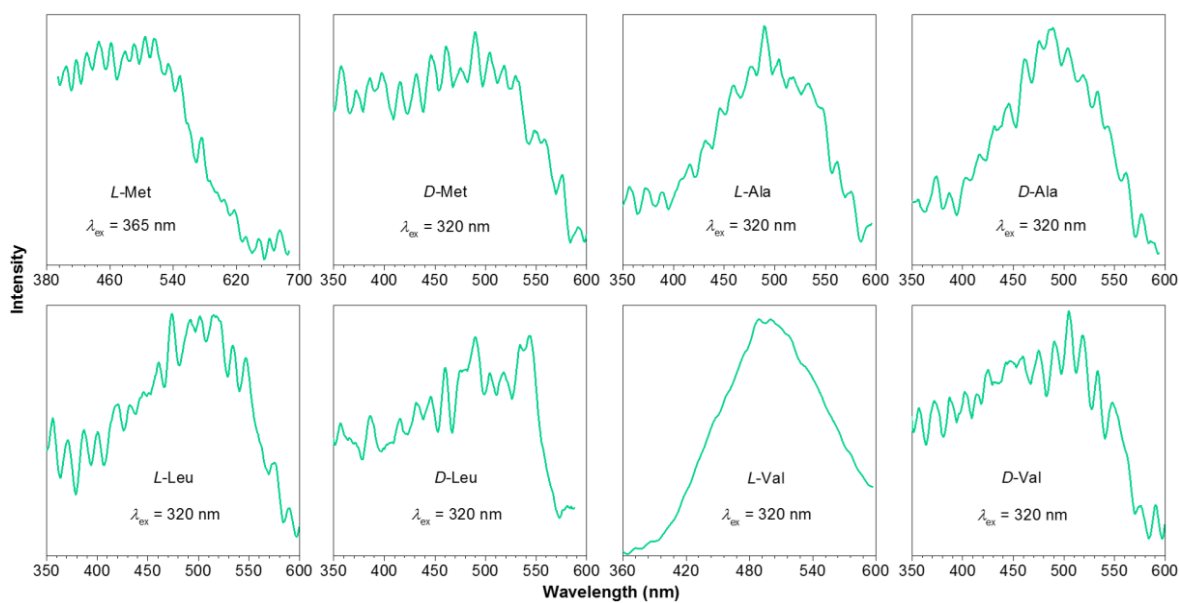

**Figure S25.** Emission spectra of recrystallized solids of *L*-Met, *D*-Met, *L*-Ala, *D*-Ala, *L*-Leu, *D*-Leu, *L*-Val, and *D*-Val with $t_d$ of 0.1 ms.

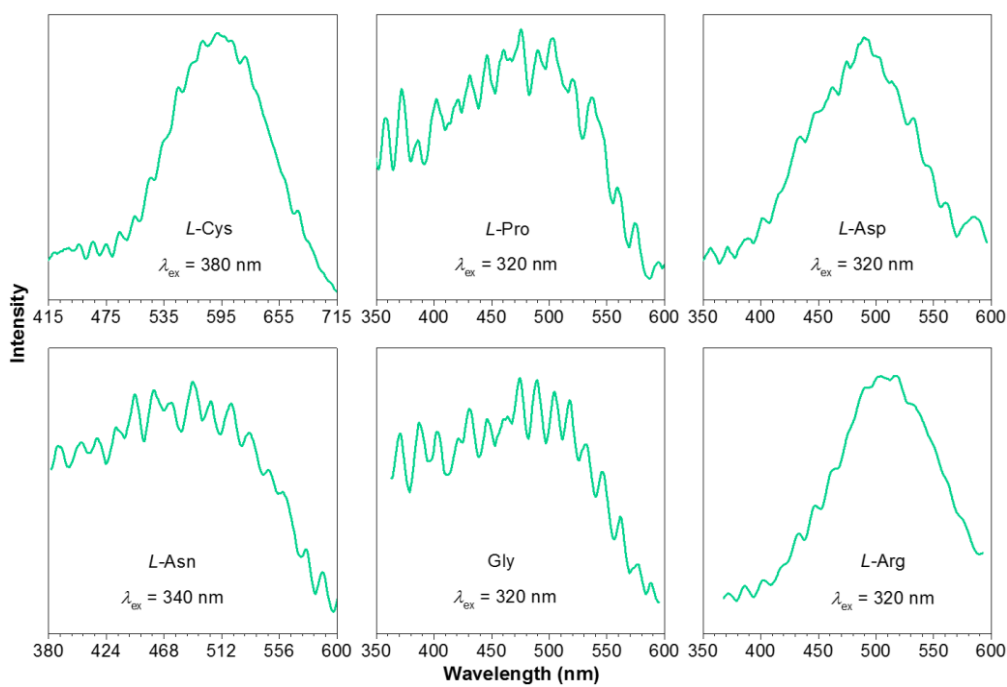

**Figure S26.** Emission spectra of recrystallized solids of *L*-Cys, *L*-Pro, *L*-Asp, *L*-Asn, Gly, and *L*-Arg with $t_d$ of 0.1 ms.

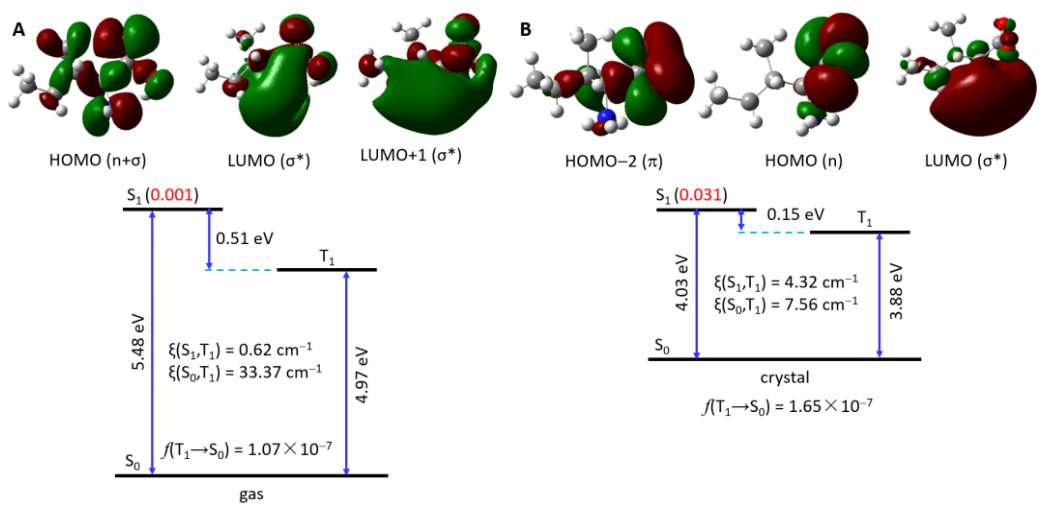

**Figure S27.** HOMO, LUMO molecular orbitals and energy levels of *L*-Ile at A) gas and B) crystal states.

As shown in **Figure S27**, from gas to crystal, the oscillator strength (*f*) of the lowest singlet state $S_1$ of *L*-Ile is increased by one order of magnitude, thus leading to bright fluorescence for *L*-Ile in crystal. It is also found that the energy gap $\Delta E_{ST}$ between $S_1$ and $T_1$ is reduced to 0.15 eV in crystal from 0.51 eV in gas, while the spin-orbit coupling (SOC) effect $\xi(S_1,T_1)$ is increased one order of magnitude (from 0.62 to 4.32 cm$^{-1}$). These two factors accelerate the intersystem crossing pathway of $S_1 \rightarrow T_1$. On the other hand, the radiative rate $k_r$ of $T_1 \rightarrow S_0$ remains unchanged according to the Einstein spontaneous emission relationship: $k_r = f \cdot E^2_{vert}/(1.499 \text{ s cm}^{-1})$, where $E_{vert}$ stands for the vertical excitation energy. In contrast, since $\xi(S_1,T_0)$ is decreased about four times, from 33.37 to 7.56 cm$^{-1}$, the nonradiative rate $k_{nr}$ of $T_1 \rightarrow S_0$ should be reduced by one order of magnitude according to $k_{nr} \propto |H_{so}|^2 \cdot FC$, where $H_{so}$ is the SOC constant and FC is the Franck-Condon factor. Consequently, RTP for *L*-Ile crystals can be detected in experimental conditions.

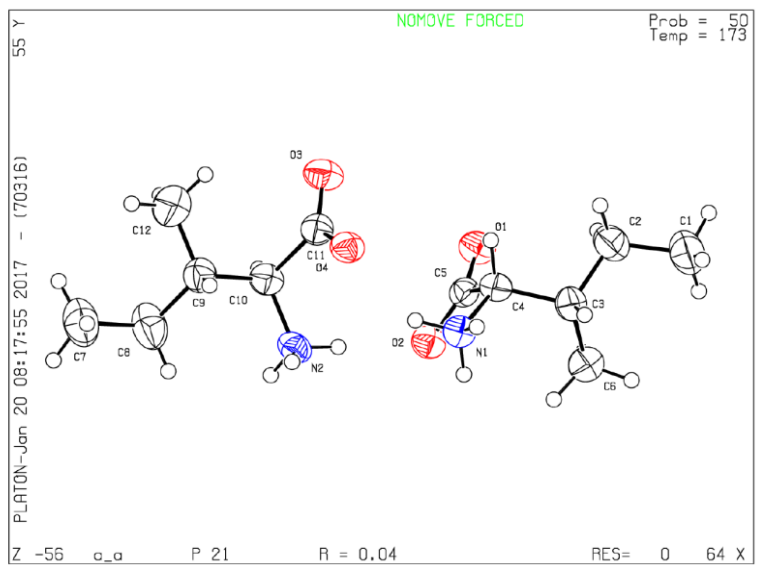

**Figure S28.** The ORTEP view of *L*-Ile with 50% ellipsoid probability.

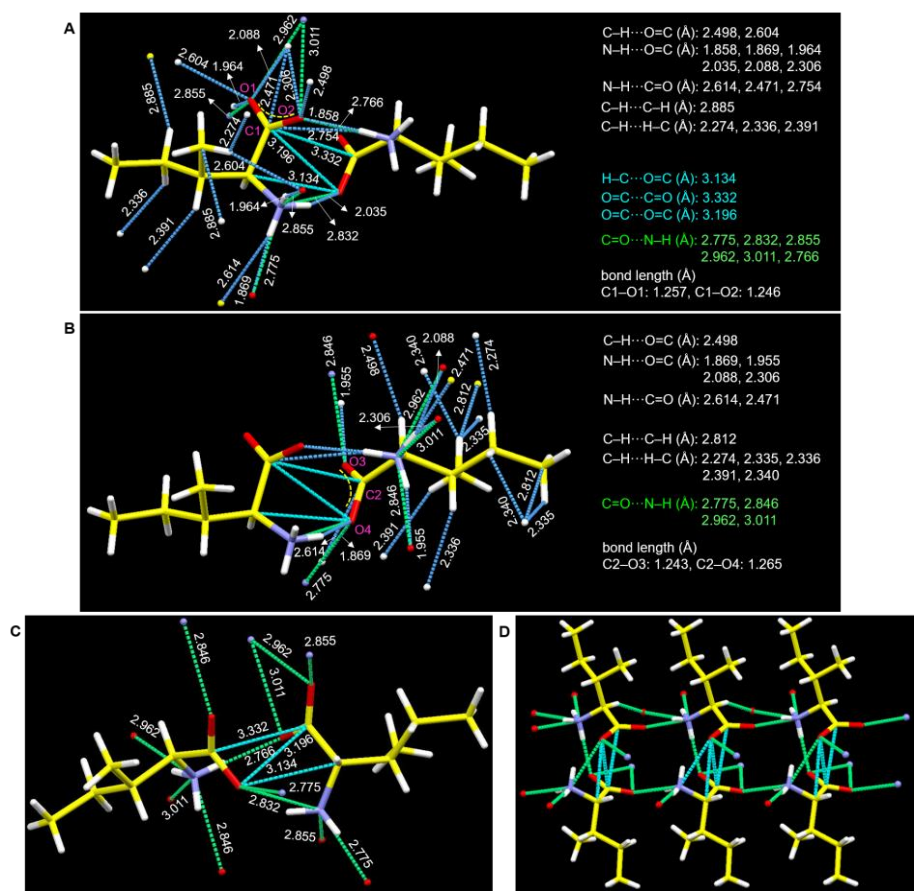

**Figure S29.** A, B) Crystal structure of *L*-Ile with denoted intermolecular interactions around two conformers. C) O=C···O=C, O=C···C=O, and N···O=C intermolecular interactions around two conformers. D) Fragmental 3D through space electronic interactions in the *L*-Ile crystals.

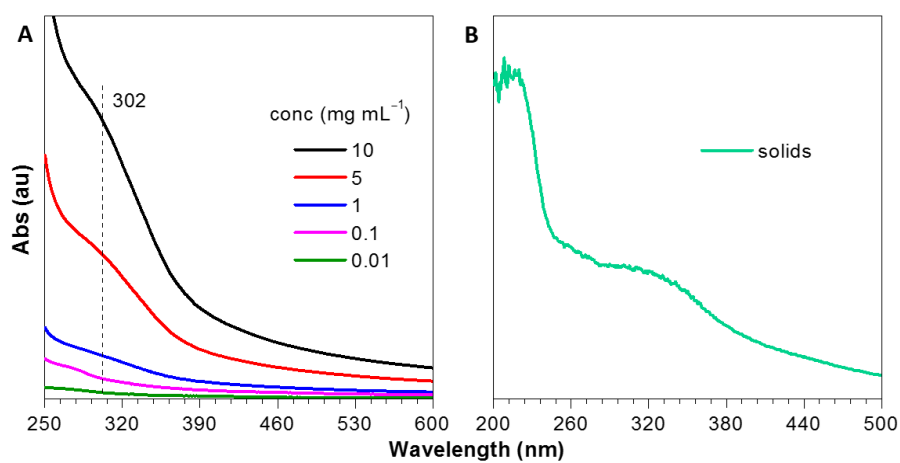

**Figure S30.** Absorption spectra of A) varying aqueous solutions and B) solid powders of *ε*-PLL.

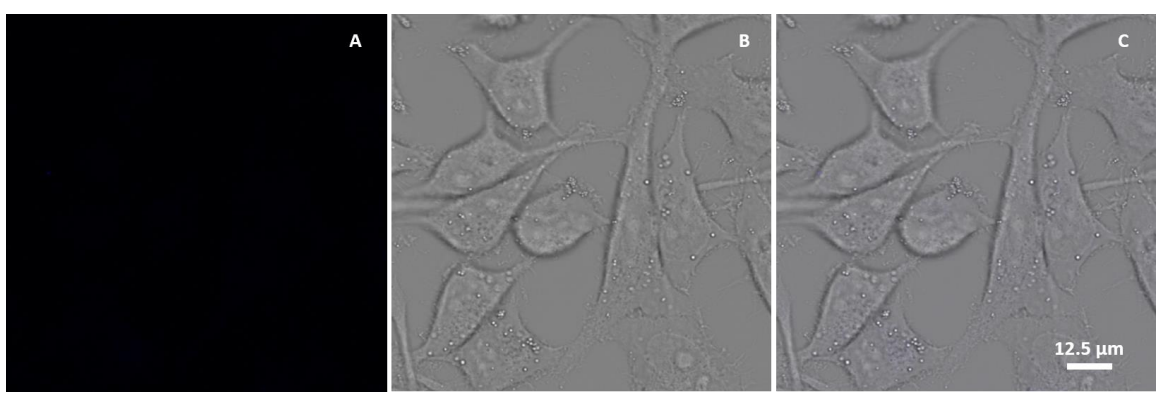

**Figure S31.** Confocal images of control HeLa cells. A) Confocal image recorded under excitation at 405 nm, B) bright field image, and C) corresponding overlayed image.